\documentclass[a4paper,fleqn]{cas-sc}

\usepackage[numbers,square,sort&compress]{natbib}
\usepackage[normalem]{ulem}
\usepackage[dvipsnames]{xcolor}
\usepackage{graphicx}
\usepackage{amsmath}
\usepackage{amsfonts}
\usepackage{amscd}
\usepackage{amssymb}
\usepackage[colorlinks]{hyperref}
\usepackage{braket}
\usepackage{enumitem}
\usepackage{multirow}
\usepackage{bm}
\usepackage{bbm}
\usepackage{units}
\usepackage{grffile}
\usepackage{mathtools}
\usepackage{xstring}
\usepackage{zref-clever}
\zcsetup{cap,abbrev}
\newcommand{\cref}[1]{\zcref{#1}}
\newcommand{\Cref}[1]{\zcref[S]{#1}}

\newcommand{\of}[1]{\omega^{#1}}
\newcommand{\bof}[1]{\bar\omega^{#1}}
\newcommand{\nmax}{{n_\text{max}}}

\newcommand{\SU}{\text{SU}}
\newcommand{\hin}{h^{f,\text{in}}}
\newcommand{\hout}{h^{f,\text{out}}}
\newcommand{\ZQCD}{Z_\text{QCD}}
\renewcommand{\ZQCD}{Z}
\renewcommand\Zbar{\bar Z}

\newcommand{\Deltatot}{\Delta}

\newcommand\aux\chi
\newcommand\pif{\int\limits_\chi}
\newcommand\Nf{N_\text{f}}
\newcommand\Nc{N_\text{c}}
\newcommand\SM{S_\text{M}}
\newcommand\SG{S^\text{G}}
\newcommand\UP{U^\text{P}}
\newcommand\DeltaP{\Delta^\text{P}}
\newcommand\bark{{\bar k}}

\graphicspath{{plots/}}
\numberwithin{equation}{section}

\newcommand{\Wg}{\widetilde{\text{W\!g}}}
\newcommand{\F}{F}
\renewcommand{\epsilon}{\varepsilon}
\DeclareMathOperator\simtwo{\dot\sim}

\newlength\sublen
\newlength\suplen
\newcommand\field[3]{
  \normalexpandarg
  \exploregroups
  \IfSubStr*{#1}{^\dagger}
  { 
    \StrDel{#1}{^\dagger}[\foo]
    \IfStrEq{#2}{}
    { 
      {\foo}^{\dagger#3}
    }
    { 
      \settowidth{\sublen}{$\scriptstyle#2$}
      \settowidth{\suplen}{$\scriptstyle\dagger$}
      {{\foo}_{#2}^\dagger}^{\hspace*{\suplen-\sublen}\,#3}
    }
  }
  { 
    {#1}_{#2}^{#3}
  }
}

\DeclareMathOperator\sgn{sgn}
\DeclareMathOperator\tr{tr}
\DeclareMathOperator{\range}{range}
\DeclareMathOperator{\subrange}{subrange}

\begin{document}

\small

\title[mode=title]{Tensor-network formulation of QCD in the strong-coupling expansion}
\author[1]{Thomas Samberger}[orcid=0009-0002-0445-386X]
\ead{thomas.samberger@ur.de}
\author[1]{Jacques Bloch}[orcid=0000-0002-8443-4804]
\ead{jacques.bloch@ur.de}
\author{Robert Lohmayer}[orcid=0000-0001-6207-4695]
\ead{robert.lohmayer@ur.de}
\author[1]{Tilo Wettig}[orcid=0000-0001-6732-9204]
\ead{tilo.wettig@ur.de}

\shorttitle{Tensor-network formulation of QCD in the strong-coupling expansion}
\shortauthors{T. Samberger, J. Bloch, R. Lohmayer, T. Wettig}

\address[1]{Institute for Theoretical Physics, University of Regensburg, 93040 Regensburg, Germany}

\date{\today}

\begin{abstract}
We present a tensor-network formulation for the strong-coupling expansion of QCD with staggered quarks at nonzero chemical potential, for arbitrary number of dimensions, colors, and flavors. We integrate out the gauge and quark degrees of freedom and rewrite the partition function as the complete trace of a tensor network. This network consists of local tensors that contain a numerical and a Grassmann part. We truncate the initial tensor at a fixed order in the inverse coupling $\beta$ and compute analytical results for the partition function, the free energy, and the chiral condensate on a $2\times2$ lattice up to order $\beta^4$. In a follow-up paper we will introduce an enhanced tensor-network method, order-separated GHOTRG, to explicitly compute the expansion coefficients of the partition function for larger lattices. To demonstrate its potential, first results obtained with this new method are already presented here.
\end{abstract}

\maketitle


\allowdisplaybreaks[4]

\section{Introduction}

The QCD phase diagram is a key research topic in modern particle physics, but its study using Monte Carlo methods in lattice QCD is hindered by the sign problem caused by the determinant of the Dirac operator, which becomes complex in the presence of a chemical potential $\mu$. Various methods developed to circumvent the sign problem, such as reweighting, Taylor expansion in $\mu$, analytic continuation from imaginary $\mu$, complex Langevin, thimbles and path optimization, have been applied to QCD, see \cite{deForcrand:2010ys,Aarts:2015tyj} for reviews. However, none of these methods can successfully reach regimes in which $\mu/T>1$, where $T$ is the temperature.  The method of dual variables shows some promise as it strongly reduces the sign problem, but until now the dualization was mainly applied to the infinite-coupling limit of QCD \cite{Rossi:1984cv,Karsch:1988zx,Fromm:2010lga,deForcrand:2009dh}. An attempt to go beyond this limit was made using the next-to-leading-order term in the strong-coupling expansion \cite{Forcrand2014} (see also \cite{Bilic:1991qy} for an early attempt in mean-field approximation). New strategies to go beyond the infinite-coupling limit in a worldline and worldsheet formulation were proposed using so-called Abelian color cycles \cite{Marchis:2017oqi} and, more recently, by introducing additional dual degrees of freedom called decoupling operator indices \cite{Gagliardi:2019cpa}. Using the method of Ref.~\cite{Gagliardi:2019cpa}, first results for the phase transition in four-dimensional QCD with one flavor of staggered quarks in the chiral limit were obtained up to order $\beta^2$ using a vertex model \cite{Unger:2025sjh}.
In most cases, the worm algorithm \cite{Prokofiev:2001zz} is used to simulate QCD in its dual formulation. 

As an alternative to Monte Carlo methods, tensor-network methods have recently been applied with success to various statistical systems. These methods can be categorized into Hamiltonian (or Hilbert-space) tensor methods and Lagrangian methods. In this paper we use the latter to study systems in thermal equilibrium and to compute the finite-temperature partition function, from which observables can be derived. Originally, the tensor renormalization group (TRG) method was proposed for two-dimensional systems \cite{Levin:2006jai}. The higher-order tensor renormalization group (HOTRG) method \cite{Xie_2012}, which is based on the higher-order singular value decomposition (HOSVD) \cite{DeLathauwer2000}, was developed to also be applicable to higher-dimensional systems. Both TRG and HOTRG are iterative procedures to compute the full contraction of a tensor network. During the blocking procedure, the bond dimension of the coarse-grid tensor (i.e., the range of the tensor indices) rapidly increases. To avoid the curse of dimensionality, the increased bond dimension is reduced using SVD or HOSVD. The two methods have been applied to a variety of problems in classical and quantum statistical physics, such as spin systems or gauge systems in two, three, and four dimensions. Even some systems with a complex action, i.e., with a sign problem, were successfully studied, for example the three-dimensional O(2) model with a chemical potential \cite{Bloch:2021mjw}.

For systems with fermions, the Hamiltonian tensor-network methods were extended to include Grass\-mann-valued tensors \cite{Gu:2010yh}. For the Lagrangian approach, which we use in this paper, the Grassmann TRG (GTRG) \cite{Shimizu:2014uva,Takeda:2014vwa} and the Grassmann HOTRG (GHOTRG) \cite{Sakai:2017jwp} methods were recently developed for cases where the Grassmann variables cannot be integrated out locally.
The local tensor contains a numerical and a Grassmann part. In GHOTRG one applies the HOTRG algorithm to the numerical part of the tensor, while the Grassmann part is self-reproducing  in each contraction step.
A comprehensive overview of the application of tensor-network methods to a range of models can be found in Ref.~\cite{Meurice:2020pxc}, where the authors provide a road map towards four-dimensional QCD.

Recently we published a first tensor-network study of two-dimensional QCD in the infinite-coupling limit with staggered quarks, where we adapted the GHOTRG method to the context of QCD \cite{Bloch:2022vqz}. We showed that the
non-local sign factors occurring in the meson-baryon-loop representation of the partition function \cite{Rossi:1984cv,Karsch:1988zx}
can be avoided by rewriting the partition function as a full contraction of a tensor network with local numerical and Grassmann tensors. The partition function is then evaluated by applying an iterative blocking procedure, which uses ideas of the original GHOTRG method but is specifically tailored for strong-coupling QCD. The method was used to compute the chiral condensate as a function of quark mass and volume, and to confirm the absence of dynamical chiral symmetry breaking in the two-dimensional case. Tensor methods are especially well-suited for this investigation, since they can be used for the very large volumes that are required for small masses.
Furthermore, the number density at nonzero chemical potential was computed, which hinted at a first-order phase transition.
This study was extended to four dimensions to investigate the phase diagram of infinite-coupling QCD, with quite promising results \cite{Milde2023}.
In \cite{Bloch:2022vqz,Milde2023} we observed that
the sign problem originating from nonzero chemical potential did not cause any numerical instabilities.

The present paper is the first in a series of papers in which we go beyond the infinite-coupling limit and consider the strong-coupling expansion of QCD up to some order $\nmax$ in $\beta\sim1/g^2$, where $g$ is the bare coupling constant. We work in $d$ Euclidean space-time dimensions, with gauge group $\SU(\Nc)$ and $\Nf$ flavors of staggered fermions. The basic idea is quite straightforward: In addition to the Taylor expansion of the Boltzmann factor involving the fermion action, as used previously in the infinite-coupling limit, one now also performs a Taylor expansion of the Boltzmann factor involving the gauge action. The gauge links (i.e., the gauge-field matrices defined on the links of the lattice) contained in the terms that arise from these expansions can still be integrated out exactly using known integrals \cite{Creutz:1978ub,Gagliardi:2018tkz,Gagliardi:2019cpa,Borisenko:2018csw}. The integrals are performed on individual color components of the gauge links and their adjoints. Consequently, the original colored Grassmann variables can be integrated out after introducing new colorless auxiliary Grassmann variables on the links of the lattice. This generates local sign factors that can be incorporated in the numerical tensor.
After integration, all color indices can be contracted, and the partition function can be rewritten as a completely contracted tensor network of local numerical and Grassmann tensors. In contrast to the infinite-coupling case, the colorless auxiliary Grassmann variables no longer correspond to baryonic degrees of freedom, but are generic fermionic degrees of freedom. Nevertheless, the GHOTRG blocking procedure remains completely identical to that of the infinite-coupling case in Ref.~\cite{Bloch:2022vqz}.

In this paper we work out the details of the derivation described above and present results for a \mbox{$2\times2$} lattice, where the expansion of the partition function, $Z=\sum_nZ_n\beta^n$, can be computed analytically to some chosen order $\nmax$.  We can then compute thermodynamical observables such as the chiral condensate in two different ways, starting from the expansion of $Z$ or of $\ln Z$.
We observe that the results obtained from the latter option show better agreement with Monte Carlo data. Furthermore, a theoretical analysis shows that an expansion of $\ln Z$ yields better results for large $\beta V$, where $V$ is the lattice volume. It would thus be preferable to use this option also on larger lattices. This requires knowledge of the $Z_n$ which, however, cannot be obtained from the standard GHOTRG method. To determine these expansion coefficients we introduce a crucial enhancement to the tensor-blocking method, which we call order-separated GHOTRG (OS-GHOTRG). The details of this enhancement will be the subject of the second paper in the series. First results for an $8\times8$ lattice obtained with this method are already shown here.

This paper is structured as follows. In \cref{sec:partition_function} we
define the partition function and Taylor-expand the exponentials of the various contributions to the action, resulting in dual variables (i.e., occupation numbers). In \cref{sec:color_indices} we present details of the enumeration of the color indices. In \cref{sec:gauge_integral_and_simplifications} we perform the gauge-link integral for a single link and simplify the result based on the color contraction with the Grassmann variables in the hopping terms. In \cref{sec:Grass_int} we integrate out the original Grassmann variables and introduce integrals over auxiliary (color- and flavorless) Grassmann variables. In \cref{sec:tensor_network} we eliminate the color degrees of freedom and construct the tensor network. In \cref{sec:prelim} we present analytical and numerical results for $2\times2$ and $8\times8$ lattices, which motivate the development of the OS-GHOTRG method. In \cref{sec:conclusions} we conclude and give an outlook to the next paper in the series.
Technical details are discussed in two appendices.

\section{Partition function and Taylor expansion}
\label{sec:partition_function}

\begin{figure}
  \centering
  \includegraphics{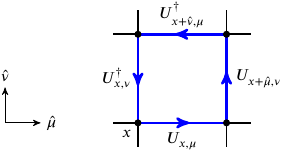}
  \caption{Visualization of the gauge links in the plaquette $\UP_{x,\mu\nu}$ defined in \eqref{eq:Plaquette_with_occupation}.}
  \label{fig:Plaq_and_adjacent_plaqs}
\end{figure}

The discretized partition function for an $\SU(\Nc)$ gauge theory with $\Nf$ flavors of staggered fermions on a $d$-dimensional Euclidean space-time lattice is given by \cite{Gattringer.2010}
\begin{equation}\label{eq:ZQCD}
  \ZQCD = \int \left[\prod_x d\psi_x d\bar\psi_x\right]\left[\prod_{x,\mu} dU_{x,\mu}\right] e^{\SM+\sum\limits_{x,\mu}(S_{x,\mu}^{+}+S_{x,\mu}^{-})+\sum\limits_{x,\mu\neq\nu} \SG_{x,\mu\nu}} \,, 
\end{equation}
where $x$ enumerates the sites of the lattice with lattice volume $V$. The different directions are denoted by $\mu\in\{1,\ldots,d\}$. All sums or products over $\mu$ or $\nu$ range from $1$ to $d$ unless different limits are stated explicitly. On every site $x=(x_1,\ldots,x_d)$ there are fermion fields $\psi_x^f$ and $\bar\psi_x^f$ ($f\in\{1,\ldots,\Nf\}$) with $\Nc$ Grassmann-valued components each, collectively denoted by $\psi_x$ and $\bar \psi_x$.  On every link $(x,\mu)$ there is a gauge-field matrix $U_{x,\mu}\in \SU(\Nc)$, which we refer to as a gauge link. We use the notation $(x,-\mu) \equiv (x-\hat\mu,\mu)$, where $\hat\mu$ is the unit vector in direction $\mu$. The Haar measure is denoted by $dU_{x,\mu}$. The Grassmann integration measure is given by
\begin{equation}
  d\psi_x d\bar\psi_x\equiv \prod_{f=1}^{\Nf}
  \prod\limits_{i=1}^{\Nc}d\psi_{x}^{f,i}d\bar\psi_{x}^{f,i}
  =\prod_{f=1}^{\Nf}\left[\coprod\limits_{i=1}^{\Nc}d\psi_{x}^{f,i} \right]
  \left[\prod_{i=1}^{\Nc}d\bar\psi_{x}^{f,i} \right],
\end{equation}
where we defined the notation
\begin{equation}\label{eq:coprod}
  \coprod_{k=a}^b\psi^k\equiv
  \begin{cases}
    \psi^b\psi^{b-1}\cdots\psi^a & \text{for }a\le b\,,\\
    1 & \text{for }a>b\\
  \end{cases}
\end{equation}
for a product of Grassmann variables in reverse order.\footnote{The symbol $\coprod$ is usually used for the coproduct, which is a very different concept. We use the symbol for notational convenience since there is no danger of confusion with the coproduct.}
The fermion action contains a mass term\footnote{\label{footnote}Note the factor of $2$ in $\SM^f$, which arises from a rescaling of the Grassmann variables by $1/\sqrt2$ to eliminate the usual factor of $1/2$ in the hopping terms \eqref{eq:hopping}.}
\begin{equation}
  \label{eq:mass}
  \SM=\sum_{f=1}^{\Nf}\SM^f \quad\text{with}\quad \SM^f=2m_f\sum_x\bar\psi_x^f\psi_x^f\,,
\end{equation}
where the $m_f$ are the quark masses and the scalar product is $\bar \psi_x^f\psi_x^f\equiv\sum_{i=1}^{\Nc}\bar\psi_x^{f,i}\psi_x^{f,i}$. The fermion action also contains the forward and backward hopping terms
\begin{equation}\label{eq:hopping}
  S_{x,\mu}^{\pm}=\sum_{f=1}^{\Nf}S_{x,\mu}^{f\pm} \quad\text{with}\quad
  \begin{cases}
   & S^{f+}_{x,\mu}=\eta_{x,\mu}e^{\mu_f\delta_{\mu,1}}\bar{\psi}_x^fU_{x,\mu}\psi_{x+\hat\mu}^f \,,\\
   & S^{f-}_{x,\mu}=-\eta_{x,\mu}e^{-\mu_f\delta_{\mu,1}}\bar{\psi}_{x+\hat\mu}^fU_{x,\mu}^\dagger\psi_x^f\,,
  \end{cases}
\end{equation}
where the hopping terms in the Euclidean time direction (for which we choose $\mu=1$) contain the quark chemical potentials $\mu_f$. The $\eta_{x,\mu}$ are the usual staggered phases $\eta_{x,\mu}=(-1)^{\sum_{\nu<\mu}x_\nu}$.
The gauge action in \eqref{eq:ZQCD} is the Wilson plaquette action with
\begin{equation}
  \label{eq:Wilson}
  \SG_{x,\mu\nu}=\frac \beta{2\Nc}\tr \UP_{x,\mu\nu}\,,
\end{equation}
where $\beta$ is the inverse coupling and the plaquette is given by the matrix product\footnote{We use the term ``plaquette'' both for the geometrical object and for the product of gauge links in the hope that no confusion will arise.}
\begin{equation}\label{eq:Plaquette_with_occupation}
  \UP_{x,\mu\nu}\equiv U_{x,\mu}U_{x+\hat\mu,\nu}U^\dagger_{x+\hat\nu,\mu}U^\dagger_{x,\nu}
\end{equation}
as illustrated in \cref{fig:Plaq_and_adjacent_plaqs}. Note that in \eqref{eq:ZQCD} we sum over oriented plaquettes, and that $\UP_{x,\nu\mu}=U_{x,\mu\nu}^{P\dagger}$.
The boundary conditions are antiperiodic in time for the Grassmann variables. All other boundary conditions are periodic.

We write the partition function as
\begin{equation}
  \ZQCD = \int \left[\prod_x d\psi_x d\bar\psi_x\right]\left[\prod_{x,\mu} dU_{x,\mu}\right]e^{\SM}\left[\prod_{x,\mu}\prod_fe^{S_{x,\mu}^{f+}}e^{S_{x,\mu}^{f-}}\right]\left[\prod_{x,\mu\neq\nu}e^{\SG_{x,\mu\nu}}\right] ,
\end{equation}
where the exponentials for the hopping terms and the gauge action are factorized over the links and plaquettes, respectively.
An intermediate goal is to integrate out the gauge links analytically. To this end, we now perform separate Taylor expansions of the exponentials for the forward and backward hopping terms on each link $(x,\mu)$ and for each flavor, with summation indices\footnote{We will frequently refer to these and other summation indices as ``occupation numbers''.} $k_{x,\mu}^f$ and $\bar k_{x,\mu}^f$, respectively. Similarly, we perform a Taylor expansion of the exponential of the gauge action on each oriented plaquette $(x,\mu\nu)$ with summation index $n_{x,\mu\nu}$. All summation indices start at zero. After performing the Taylor expansions we rewrite the products of sums as sums of products,
\begin{equation}
  \label{eq:Ztaylor}
  \ZQCD = \int \left[\prod_x d\psi_x d\bar\psi_x\right]\left[\prod_{x,\mu} dU_{x,\mu}\right]
  e^{\SM} \sum_{\bm{k}, \bm{n}} 
  \left[\prod_{x,\mu}\prod_f\frac{(S_{x,\mu}^{f+})^{k_{x,\mu}^f}(S_{x,\mu}^{f-})^{\bar k_{x,\mu}^f}}{k_{x,\mu}^f!\bar k_{x,\mu}^f!}\right]
  \left[\prod_{x,\mu\neq\nu} \frac{(\SG_{x,\mu\nu})^{n_{x,\mu\nu}}}{n_{x,\mu\nu}!}\right] 
\end{equation}
with tuples $\bm{k}\equiv((k_{x,\mu}^f,\bar{k}_{x,\mu}^f)| \, \forall \, x,\mu,f)$ and $\bm{n}\equiv(n_{x,\mu\nu}| \, \forall \, x,\mu,\nu;\mu\neq\nu)$ that contain all link and plaquette occupation numbers of the lattice. These occupation numbers can be viewed as dual fields, which also have periodic boundary conditions.

To be able to express the partition function as the trace of a tensor network, we exchange the order of integration and summation to obtain
\begin{equation}
  \label{eq:Z2}
  \ZQCD = \sum_{\bm{k}, \bm{n}} \int \left[\prod_x d\psi_x d\bar\psi_x\right]\left[\prod_{x,\mu} dU_{x,\mu}\right]
  e^{\SM}
  \left[\prod_{x,\mu}\prod_f\frac{(S_{x,\mu}^{f+})^{k_{x,\mu}^f}(S_{x,\mu}^{f-})^{\bar k_{x,\mu}^f}}{k_{x,\mu}^f!\bar k_{x,\mu}^f!}\right]
  \Biggl[\prod_{x,\mu\neq\nu} \frac{(\SG_{x,\mu\nu})^{n_{x,\mu\nu}}}{n_{x,\mu\nu}!}\Biggl] .
\end{equation}
For the summation over $\bm{k}$ this exchange is justified since the sums are finite due to the nature of the Grassmann variables and since the gauge integral is convergent.  However, because the summation over $\bm{n}$ involves infinite sums the exchange is potentially questionable. A formal convergence analysis appears to be nontrivial, and
therefore this issue needs to be kept in mind when numerical results are interpreted.

In \cref{sec:gauge_integral,sec:Grass_int} we will first compute the gauge integral on every link and then integrate over the Grassmann fields on every site, respectively. This needs some preparatory steps, which we discuss next.

\section{Disentangling color indices}
\label{sec:color_indices}

To do the gauge-link integrals we have to rearrange the gauge links coming from the hopping terms and the plaquette action in such a way that all contributions from a specific gauge link are gathered. This is done by writing all summations over color indices explicitly, which allows us to move components of the gauge links freely. We use the convention that gauge links $U$ always carry color indices $i$ and $j$, while adjoint gauge links $U^\dagger$ carry color indices $\ell$ and $m$. To distinguish all the color indices occurring in \eqref{eq:Z2} we need to assign two additional labels to each color index. First, in order to distinguish the color indices of different gauge links, we assign a link label to the color indices. Second, as the same gauge link $U_{x,\mu}$ and its adjoint $U_{x,\mu}^\dagger$ can occur several times in each term contributing to the partition function, we distinguish the color indices corresponding to these different occurrences by a label $a$. Hence, our color indices are $i_{x,\mu}^a$, $j_{x,\mu}^a$, $\ell_{x,\mu}^a$, and $m_{x,\mu}^a$, again with periodic boundary conditions.

In the following we discuss the label $a$ in detail. Some of the ensuing formulas may look somewhat complicated, but this is just a matter of enumeration that can be skipped over at a first reading.
The range of the label $a$ consists of several subranges.  For $U_{x,\mu}$, the subrange of $a$ from the forward hopping term for flavor $f$ contains $k_{x,\mu}^f$ elements, and there are $\Nf$ such subranges. For the plaquette terms, we note that in $d$ dimensions, a given gauge link $U_{x,\mu}$ is contained in $2(d-1)$ plaquettes, i.e., $\UP_{x,\mu\nu}$ and $\UP_{x-\hat\nu,\nu\mu}$ with $\nu\ne\mu$. Hence there are $2(d-1)$ subranges of $a$ for the plaquette terms. In summary, the sizes of the $\Nf+2(d-1)$ subranges of $a$ for the various contributions are given by
\begin{subequations} \label{colorindices}
  \begin{align}
    U_{x,\mu}: \quad |\subrange(a)| &=
    \begin{cases}
      k_{x,\mu}^f & \text{forward fermion hopping for flavor $f$}\,, \\
      n_{x,\mu\nu} & \text{plaquettes $\UP_{x,\mu\nu}$ for $\nu\ne\mu$}\,, \\
      n_{x-\hat\nu,\nu\mu}  & \text{plaquettes $\UP_{x-\hat\nu,\nu\mu}$ for $\nu\ne\mu$}\,,
    \end{cases}
    \intertext{where $|\cdots|$ denotes the number of elements of the (sub)range. Similarly, for $U_{x,\mu}^\dagger$ the subrange sizes are}
    U_{x,\mu}^\dagger : \quad |\subrange(a)| &=
    \begin{cases}
      \bar k_{x,\mu}^f & \text{backward fermion hopping for flavor $f$}\,, \\
      n_{x,\nu\mu} & \text{plaquettes $\UP_{x,\nu\mu}$ for $\nu\ne\mu$}\,, \\
      n_{x-\hat\nu,\mu\nu}  & \text{plaquettes $\UP_{x-\hat\nu,\mu\nu}$ for $\nu\ne\mu$}\,.
    \end{cases}
  \end{align}
\end{subequations}
We now define
\begin{equation}
  \label{eq:k_kbar}
  k_{x,\mu}\equiv\sum_fk_{x,\mu}^f \quad\text{and}\quad \bar k_{x,\mu}\equiv\sum_f\bar k_{x,\mu}^f
\end{equation}
and combine the subranges of $a$ to a full range, whose size is given by
\begin{subequations}\label{eq:g}
  \begin{align}
    U_{x,\mu}: \quad |\range(a)|&=g_{x,\mu}\equiv k_{x,\mu}+\sum_{\nu\atop{\nu\ne\mu}}(n_{x,\mu\nu}+n_{x-\hat\nu,\nu\mu})\,,\\
    U_{x,\mu}^\dagger: \quad |\range(a)|&=\bar g_{x,\mu}\equiv \bar k_{x,\mu}+\sum_{\nu\atop{\nu\ne\mu}}(n_{x,\nu\mu}+n_{x-\hat\nu,\mu\nu})\,,
  \end{align}
\end{subequations}
respectively.
To access the different subranges within the full range, we define the offsets\footnote{The offsets $\omega$ and $\bar\omega$ correspond to $U$ and $U^\dagger$, respectively. The superscript $\pm\nu$ indicates that the plaquette of interest extends in the positive/negative $\nu$-direction perpendicular to the link $(x,\mu)$.}
\begin{subequations}
  \begin{alignat}{2}
    \label{eq:kappa}
    \kappa_{x,\mu}^f&\equiv\sum_{f'=1}^{f-1}k_{x,\mu}^{f'}\,, &\qquad\quad
    \bar\kappa_{x,\mu}^f&\equiv\sum_{f'=1}^{f-1}\bar k_{x,\mu}^{f'}\,,\\
    \of{+\nu}_{x,\mu} &\equiv k_{x,\mu} + \sum_{{\sigma<\nu}\atop{\sigma\neq\mu}} (n_{x,\mu\sigma} + n_{x-\hat\sigma,\sigma\mu})\,, &\qquad\quad
    \of{-\nu}_{x,\mu} &\equiv \of{+\nu}_{x,\mu} + n_{x,\mu\nu}\,, \\
    \bof{+\nu}_{x,\mu} &\equiv \bar k_{x,\mu} + \sum_{{\sigma<\nu}\atop{\sigma\neq\mu}} (n_{x,\sigma\mu} + n_{x-\hat\sigma,\mu\sigma})\,, &
    \bof{-\nu}_{x,\mu} &\equiv \bof{+\nu}_{x,\mu} + n_{x,\nu\mu}\,.
  \end{alignat}
\end{subequations}
With these definitions and using the Einstein summation convention, the color contractions in the terms
coming from $S^{f+}_{x,\mu}$, $S^{f-}_{x,\mu}$, and $\SG_{x,\mu\nu}$ are written as
\begin{subequations}\label{eq:color_ind_Wilson}
\begin{align}
  (\bar\psi_x^fU_{x,\mu}\psi_{x+\hat\mu}^f)^{k_{x,\mu}^f}
  &=\prod_{a=1}^{k_{x,\mu}^f}
    \field{{\bar\psi}}{x}{f,i^{a+\kappa_{x,\mu}^f}_{x,\mu}}
    \field{U}{x,\mu}{i^{a+\kappa_{x,\mu}^f}_{x,\mu},j^{a+\kappa_{x,\mu}^f}_{x,\mu}}
    \field{\psi}{x+\hat\mu}{f,j_{x,\mu}^{a+\kappa_{x,\mu}^f}} \,,\label{eq:color_ind_Wilson_a}\\
  (\bar{\psi}_{x+\hat\mu}^fU_{x,\mu}^\dagger\psi_x^f)^{\bar k_{x,\mu}^f}
  &=\prod_{a=1}^{\bar k_{x,\mu}^f}
    \field{{\bar{\psi}}}{x+\hat\mu}{f,\ell^{a+\bar\kappa_{x,\mu}^f}_{x,\mu}}
    \field{U^\dagger}{x,\mu}{\ell^{a+\bar\kappa_{x,\mu}^f}_{x,\mu},m^{a+\bar\kappa_{x,\mu}^f}_{x,\mu}}
    \field{\psi}{x}{f,m_{x,\mu}^{a+\bar\kappa_{x,\mu}^f}}
    \,,\label{eq:color_ind_Wilson_b}\\
  \left(\tr \UP_{x,\mu\nu}\right)^{n_{x,\mu\nu}}
  &= \prod_{a=1}^{n_{x,\mu\nu}} \Delta_{x,\mu\nu}^a
    \field{U}{x,\mu}{i_{x,\mu}^{a+\of{+\nu}_{x,\mu}},j_{x,\mu}^{a+\of{+\nu}_{x,\mu}}}
    \field{U}{x+\hat\mu,\nu}{i_{x+\hat\mu,\nu}^{a+\of{-\mu}_{x+\hat\mu,\nu}},j_{x+\hat\mu,\nu}^{a+\of{-\mu}_{x+\hat\mu,\nu}}}
    \field{U^\dagger}{x+\hat\nu,\mu}{\ell_{x+\hat\nu,\mu}^{a+\bof{-\nu}_{x+\hat\nu,\mu}},m_{x+\hat\nu,\mu}^{a+\bof{-\nu}_{x+\hat\nu,\mu}}}
    \field{U^\dagger}{x,\nu}{\ell_{x,\nu}^{a+\bof{+\mu}_{x,\nu}},m_{x,\nu}^{a+\bof{+\mu}_{x,\nu}}}\,,
    \label{eq:trP}
\end{align}
\end{subequations}
where we have rewritten the matrix multiplications and traces in \eqref{eq:trP} using the quantity
\begin{equation}\label{eq:Delta_a}
  \Delta_{x,\mu\nu}^a \equiv
  \delta_{j_{x,\mu}^{a+\of{+\nu}_{x,\mu}},i_{x+\hat\mu,\nu}^{a+\of{-\mu}_{x+\hat\mu,\nu}}}
  \delta_{j_{x+\hat\mu,\nu}^{a+\of{-\mu}_{x+\hat\mu,\nu}},\ell_{x+\hat\nu,\mu}^{a+\bof{-\nu}_{x+\hat\nu,\mu}}}
  \delta_{m_{x+\hat\nu,\mu}^{a+\bof{-\nu}_{x+\hat\nu,\mu}},\ell_{x,\nu}^{a+\bof{+\mu}_{x,\nu}}}
  \delta_{m_{x,\nu}^{a+\bof{+\mu}_{x,\nu}},i_{x,\mu}^{a+\of{+\nu}_{x,\mu}}}\,.
\end{equation}
The motivation behind this rewriting is that in the construction of the tensor formulation below it is more convenient to retain independent indices on $U$ and $U^\dagger$ to be able to manipulate individual indices.
We see from \eqref{eq:color_ind_Wilson} that the color indices $i_{x,\mu}^a$ and $m_{x,\mu}^a$ can be associated with site $x$, while $j_{x,\mu}^a$ and $\ell_{x,\mu}^a$ can be associated with site $x+\hat\mu$.

We can now gather all gauge-field contributions corresponding to the same link. After all the dust has settled, the partition function can be written as
\begin{equation}\label{eq:part_func1}
  \ZQCD = \sum_{\bm{k}, \bm{n}} 
  \int \left[\prod_x d\psi_x d\bar\psi_x\right]
  \mathcal{G}
  \mathcal{W}
  \Deltatot
  \,
  \prod_{x,\mu}
  \mathcal{I}^{g_{x,\mu}\bar g_{x,\mu}}\,,
\end{equation}
where all contributions involving the same gauge-field matrix $U$ are combined in the integral
\begin{equation}\label{eq:I}
  (\mathcal{I}^{g\bar g})^{\bm{i}\bm{j},\bm{\ell}\bm{m}}
  \equiv
  \int_{\SU(\Nc)}dU\left[\prod_{a=1}^{g}\field{U}{}{i^{a}j^{a}}\right]\left[\prod_{a=1}^{\bar g}\field{U^\dagger}{}{\ell^{a}m^{a}}\right]
\end{equation}
with color indices
\begin{equation}\label{eq:i,j,l,m}
  \bm{i}\equiv(i^1,\dots,i^{g})\,,\quad \bm{j}\equiv(j^1,\dots,j^{g})\,,\quad
  \bm{\ell}\equiv(\ell^1,\dots,\ell^{{\bar g}}) \,,\quad
  \bm{m}\equiv(m^1,\dots,m^{{\bar g}})\,. 
\end{equation}
The gauge-link occupation numbers $g$ and $\bar g$ in \eqref{eq:part_func1} are defined in \eqref{eq:g}.
All Grassmann variables on the lattice are combined in the factor
\begin{equation}
  \label{eq:Grassmann_block}
  \mathcal{G}=e^{\SM}\prod_{x,\mu}\mathcal{G}_{x,\mu}
  \quad\text{with}\quad
  \mathcal{G}_{x,\mu}=\prod_f
  \left[\prod_{a=1}^{k_{x,\mu}^f}
  \field{\bar{\psi}}{x}{f,i_{x,\mu}^{a+\kappa_{x,\mu}^f}}
  \field{\psi}{x+\hat\mu}{f,j_{x,\mu}^{a+\kappa_{x,\mu}^f}}\right]
  \left[\prod_{a=1}^{\bar k_{x,\mu}^f}
  \field{\psi}{x}{f,m_{x,\mu}^{a+\bar\kappa_{x,\mu}^f}}
  \field{\bar{\psi}}{x+\hat\mu}{f,\ell_{x,\mu}^{a+\bar\kappa_{x,\mu}^f}}\right] ,
\end{equation}
where in the last factor we have changed the order of the Grassmann variables to eliminate the minus sign in \eqref{eq:hopping}.  The integration of $\mathcal{G}$ will be discussed in \cref{sec:Grass_int}.  The remaining quantities in \eqref{eq:part_func1} are
\begin{align} \label{Delta}
  \Deltatot &\equiv \prod_{x,\mu\ne\nu}\prod_{a=1}^{n_{x,\mu\nu}} \Delta_{x,\mu\nu}^a\,,\\
    \label{numpref}
  \mathcal{W}
  &\equiv \left[\prod_{x,\mu\ne\nu} \frac{{\left(\beta/2\Nc\right)}^{n_{x,\mu\nu}}}{n_{x,\mu\nu}!}\right]
    \left[\prod_{x,\mu}\prod_f
    \frac{\eta_{x,\mu}^{k_{x,\mu}^f+\bar k_{x,\mu}^f}}{k_{x,\mu}^f!\bar k_{x,\mu}^f!}\right]
    \left[\prod_x\prod_fe^{\mu_f(k_{x,1}^f-\bar k_{x,1}^f)}\right].
\end{align}

\section{Gauge-link integral and simplifications}
\label{sec:gauge_integral_and_simplifications}

\subsection{Gauge-link integral}
\label{sec:gauge_integral}

Now that we have collected all gauge-field contributions for every link, we are ready to evaluate the resulting $\SU(\Nc)$ integrals. The relevant integral, given in \eqref{eq:I}, was first solved by Creutz \cite{Creutz:1978ub}, but we will use a more efficient solution, which was recently presented by Gagliardi and Unger~\cite{Gagliardi:2018tkz,Gagliardi:2019cpa} and by Borisenko et al.~\cite{Borisenko:2018csw}. These authors gave expressions for the case of $g\ge\bar g$ and pointed out that
using the defining property of the Haar measure one obtains $(\mathcal{I}^{g\bar g})^{\bm{ij},\bm{\ell m}}=(\mathcal{I}^{\bar g g})^{\bm{\ell m},\bm{ij}}$, from which the case of $g<\bar g$ follows. After defining $p\equiv\min(g,\bar g)$ and $q\equiv|g-\bar g|/\Nc$, both cases can be combined in the result
\begin{equation}\label{eq:gauge_int}
  (\mathcal{I}^{g\bar g})^{\bm{ij},\bm{\ell m}} =
  \left[\prod_{s=1}^{\Nc-1}\frac{s!}{(s+q)!}\right] 
  \sum_{\pi,\sigma\in S_{p}}\Wg_{\Nc}^{q,p}(\pi\cdot\sigma^{-1})
  \sum_{\alpha\in A}
  (I^{\text{b}}_{\alpha\pi\sigma})^{\bm{i}\bm{m}}
  (I^\text{e}_{\alpha\pi\sigma})^{\bm{j}\bm{\ell}}
  \qquad \text{for } q \in\mathbb{N}_0 \,.
\end{equation}
If $q\notin\mathbb{N}_0$ the integral is zero.
In the remainder of this section we describe the various ingredients of the result \eqref{eq:gauge_int}.

The sums over $\pi$ and $\sigma$ run over all elements of the symmetric group $S_p$. Here, we consider the special case that a permutation $\pi$ acts on the sequence of numbers $(1,\dots,p)$, resulting in the permuted sequence $(\pi_1,\dots,\pi_p)$, and similarly for $\sigma$. Note that we use the same symbol $\pi$ (or $\sigma$) for both the permutation and the permuted sequence.

The generalized Weingarten function $\Wg$ is defined for $\pi\in S_p$ by \cite{Gagliardi:2018tkz}
\begin{equation}\label{eq:genWeingarten}
  \Wg_{\Nc}^{q,p}(\pi) \equiv \frac{1}{(p!)^2}\sum_{{\lambda\atop|\lambda|\le \Nc}}\frac{d_\lambda^2}{s_{\lambda}^{(\Nc+q)}}\chi^\lambda(\pi)\,.
\end{equation}
Here, $\lambda$ stands both for an irreducible representation (irrep) of $S_p$ and for the corresponding partition of $p$.\footnote{A partition $\lambda$ of the integer $p$ is a sequence $(\lambda_1,\dots,\lambda_k)$ of integers such that $\lambda_i\ge\lambda_{i+1}>0$ and $\sum_{i=1}^k\lambda_i=p$. The length $|\lambda|$ of the partition is the number of elements in the sequence, i.e., $|\lambda|=k$.}
The sum over $\lambda$ is restricted to partitions with length $|\lambda|\le \Nc$. The dimension of the irrep $\lambda$ is denoted by $d_\lambda$, and $\chi^\lambda(\pi)$ is the character of $\pi$ in irrep $\lambda$.
The quantity $s_\lambda^{(\Nc+q)}$ is defined by
\begin{equation}
  s_{\lambda}^{(\Nc+q)}\equiv\prod_{1\leq i<j\leq \Nc+q}\frac{\lambda_i-\lambda_j+j-i}{j-i}\,,
\end{equation}
which is the Schur polynomial evaluated at the $(\Nc + q)$-dimensional point $(1, 1, \dots, 1)$.
	
The definitions of $p$ and $q$ imply $\max(g,\bar g)=q\Nc+p$.
The sum over $\alpha\in A$ in \eqref{eq:gauge_int} runs over all possible ways to pick $q$ subsets of size $\Nc$ each from the set $\{1,\dots,q\Nc+p\}$ in such a way that all numbers that have been picked are distinct. There are $\frac{(q\Nc+p)!}{p!q!(\Nc!)^{q}}$ ways to do so. The remaining subset has size $p$. For later use, we order the numbers in each of these $q+1$ subsets in increasing order, i.e., the subsets become increasing sequences.
Therefore, a particular grouping $\alpha$ can be written using integers $\alpha^c_s$ ($1\le s\le q$, $1\le c\le \Nc$) as a set of increasing sequences,
\begin{subequations}\label{eq:alpha_gamma}
  \begin{gather}\label{eq:alpha}
    \alpha\equiv \left\{(\alpha^1_1, \dots, \alpha_1^{\Nc}),\dots ,(\alpha_q^1, \dots, \alpha_q^{\Nc})\right\} \quad
    \text{with } \alpha^c_s < \alpha^{c+1}_{s}\,.
    \intertext{The remaining increasing sequence is denoted by $\gamma(\alpha)$. It is completely fixed by $\alpha$ and is written using integers $\gamma_s$ ($1\le s\le p$) as}
    \gamma(\alpha)\equiv\left(\gamma_1, \ldots, \gamma_p\right) \quad\text{with }\gamma_s<\gamma_{s+1}\,.
  \end{gather}
\end{subequations}
Recall that the $\alpha_s^c$ and $\gamma_s$ in \eqref{eq:alpha_gamma} are all distinct and from $\{1,\dots,q\Nc+p\}$.

All color indices in \eqref{eq:gauge_int} are collected in the expressions%
\footnote{As mentioned after \eqref{eq:Delta_a}, the color indices $i$ and $m$ always correspond to the site at the beginning of the link, while $j$ and $\ell$ correspond to the site at the end of the link.}
\begin{align}\label{eq:Ibf}
  (I^\text{b}_{\alpha\pi\sigma})^{\bm{i}\bm{m}}\equiv
  \begin{cases}
    \varepsilon_{\bm i_{\alpha}}^{\otimes q}\delta^{\bm m_\pi}_{\bm i_{\gamma}}&\text{for~}g\geq \bar g\,, \\
    \varepsilon^{\otimes q}_{\bm m_{\alpha}}\delta_{\bm m_{\gamma}}^{\bm i_\sigma} &\text{for}~  g< \bar g\,,
  \end{cases}
  \qquad\text{and}\qquad
  (I^\text{e}_{\alpha\pi\sigma})^{\bm{j}\bm{\ell}}\equiv
  \begin{cases}
    \varepsilon_{\bm j_{\alpha}}^{\otimes q}\delta^{\bm \ell_\sigma}_{\bm j_{\gamma}}&\text{for~}g\geq \bar g\,, \\
    \varepsilon^{\otimes q}_{\bm \ell_{\alpha}}\delta_{\bm \ell_{\gamma}}^{\bm j_\pi} &\text{for}~  g< \bar g\,,
  \end{cases}
\end{align}
where b and e stand for the beginning and the end of the link, respectively, and where we omitted the $\alpha$-dependence of $\gamma$ for readability.
The $q$-fold Levi-Civita tensor and the Kronecker delta of two tuples are defined as
  \begin{equation}\label{eq:def_qfold_Levi_and_Delta}
  \quad\epsilon^{\otimes q}_{\bm i_\alpha} \equiv \prod_{s=1}^q \epsilon_{i^{\alpha_s^1}, \ldots, i^{\alpha_s^{\Nc}}}
  \qquad\text{and}\qquad
  \delta^{\bm m_\pi}_{\bm i_\gamma}  \equiv \prod_{s=1}^p \delta_{m^{\pi_s}, i^{\gamma_s}} \,,
\end{equation}
respectively.
From \eqref{eq:Ibf} and \eqref{eq:def_qfold_Levi_and_Delta} we see that the elements of $\alpha$, $\gamma$, $\pi$, and $\sigma$ correspond to the additional label $a$ on the color indices we defined at the beginning of \cref{sec:color_indices}.

Note that in Ref.~\cite{Gagliardi:2019cpa} the sums in \eqref{eq:gauge_int} are eventually replaced by a sum over so-called decoupling operator indices, which arise from explicitly expressing the character of $\pi\cdot\sigma^{-1} \in S_p$, see \eqref{eq:gauge_int} and \eqref{eq:genWeingarten}, in terms of individual matrix elements.
The motivation is that the sum over permutations $\pi$ and $\sigma$ generates oscillating signs in the Weingarten functions, which leads to a sign problem in the Monte Carlo simulation of their dual formulation \cite{Gagliardi:2018tkz}. These signs will not be a problem in the tensor formulation, which we develop here, and therefore we keep the sum over permutations.

\subsection{Range reduction due to Grassmann variables}

Recall that the integral \eqref{eq:gauge_int} comes from a link between two sites.
In the partition function, all color indices in $(I_{\alpha\pi\sigma}^\text{b})^{\bm{i}\bm{m}}$ and $(I^\text{e}_{\alpha\pi\sigma})^{\bm{j}\bm\ell}$ that originate from Grassmann hopping terms are contracted with the color indices of the Grassmann variables at these two sites.
In this subsection we will show that this observation leads to a simplification of the sum over $\alpha$, $\pi$, and $\sigma$. In Ref.~\cite{Gagliardi:2019cpa} a similar reduction was briefly mentioned for the sum over the decoupling operator indices.

We first derive a useful intermediate result.
Consider a function $f$ depending on color indices $i^1,\dots,i^k$ that are contracted with Grassmann variables $\chi^{i^1},\dots,\chi^{i^k}$. Also consider a permutation $\rho\in S_k$.  Using the Einstein summation convention, we have 
\begin{equation}\label{eq:property_grass_perm_weing}
  f(i^{\rho_1},\cdots,i^{\rho_k})\prod_{b=1}^{k}\chi^{i^{b}}
  =f\big(i^{\rho_1},\cdots,i^{\rho_k})\sgn(\rho)\prod_{b=1}^{k}\chi^{i^{\rho_b}}
  =\sgn(\rho)f(i^1,\cdots,i^k)\prod_{b=1}^{k}\chi^{i^b}\,.
\end{equation}
In the first step, we reordered the Grassmann variables, which produces a prefactor given by the sign of the permutation $\rho$.
In the second step, the equality follows from renaming the summation variables.

We will use \eqref{eq:property_grass_perm_weing} with $\chi$ replaced by a single color vector of Grassmann variables $\psi^f_x$ or $\bar\psi^f_x$, and for color indices in the functions $I_\text{b}$ and $I_\text{e}$ in \eqref{eq:Ibf} that originate from Grassmann hopping terms. We thus need to identify these color indices in our enumeration scheme.
In \eqref{eq:Ibf} we see that for $g\ge \bar g$, the entries of $\alpha$ and $\gamma$ label the indices $i$ and $j$ of $U$, while the entries of $\pi$ and $\sigma$ label the indices $\ell$ and $m$ of $U^\dagger$. For $g<\bar g$ it is the other way around. If we define
\begin{equation}
  \varkappa^f \equiv
  \begin{cases}
    \kappa^f &\text{for~} g \geq \bar g\,, \\
    \bar\kappa^f &\text{for}~  g < \bar g\,,
  \end{cases}
  \qquad
  \bar \varkappa^f \equiv
  \begin{cases}
    \bar \kappa^f &\text{for~} g \geq \bar g\,, \\
    \kappa^f &\text{for}~  g < \bar g
  \end{cases} 
\end{equation}
with $\kappa^f$ and $\bar\kappa^f$ defined in \eqref{eq:kappa}, then for given $\alpha$, $\pi$, and $\sigma$, the entries coming from the forward and backward hopping terms of flavor $f$ correspond to
\begin{equation}
  \varkappa^f<\alpha_s^c,\gamma_s\le\varkappa^{f+1} \quad\text{and}\quad \bar\varkappa^f<\pi_s,\sigma_s\le\bar\varkappa^{f+1}\,,
\end{equation}
respectively. For given $f$, there are $\varkappa^{f+1}-\varkappa^f$ entries satisfying the first inequality. The number of entries of $\gamma$ for which $\varkappa^f<\gamma_s\le\varkappa^{f+1}$ will be called $z^f$. This implies that there are $\varkappa^{f+1}-\varkappa^f-z^f$ entries with $\varkappa^f<\alpha_s^c\le\varkappa^{f+1}$. Note that $\varkappa^{f+1}-\varkappa^f$ equals $k^f$ for $g\ge\bar g$ and $\bar k^f$ for $g<\bar g$, respectively.

It will turn out to be useful to introduce two different equivalence relations, denoted by $\sim$ and $\simtwo$. We define two groupings $\alpha,\alpha'\in A$ to be equivalent (written as $\alpha\sim\alpha'$) if the pairs $\alpha,\gamma(\alpha)$ and $\alpha',\gamma(\alpha')$ can be transformed into each other by permuting, for every flavor $f$ separately, the entries $\alpha_s^c$ and $\gamma_s$ satisfying $\varkappa^f<\alpha_s^c,\gamma_s\le\varkappa^{f+1}$.
Furthermore, we define two permutations $\pi,\pi'\in S_p$ to be equivalent (written as $\pi\simtwo\pi'$) if the corresponding sequences can be transformed into each other by permuting the numbers $\pi_s$ with $\bar\varkappa^f<\pi_s\le\bar\varkappa^{f+1}$ (for every flavor $f$ separately)
and by permuting the numbers $s$ with $\sum_{f'=1}^{f-1}z^{f'}<s\le\sum_{f'=1}^fz^{f'}$ (for every flavor $f$ separately).\footnote{Note that the numbers $\pi_s$ appear in \eqref{eq:def_qfold_Levi_and_Delta} in Kronecker deltas together with $\gamma_s$. Since the numbers $\gamma_s$ are in increasing order, the $z^f$ entries of $\gamma$ originating from the Grassmann hopping terms of flavor $f$ start at $s=\sum_{f'=1}^{f-1}z^{f'}$. Since we cannot permute the entries of $\gamma$, we permute the entries of the sequence $\pi$ with the same $s$-indices.} With these two definitions we can use \eqref{eq:property_grass_perm_weing} to simplify the sum over $\alpha$, $\pi$, and $\sigma$. To this end we write
\begin{equation}\label{eq:aux_range_splitting}
  \sum_{\alpha\in A}=\sum_{\alpha\in A^\sim}\,\,\,\sum_{\alpha' \in[\alpha]}
  \quad\text{and}\quad
  \sum_{\pi\in S_{p}}=\sum_{\pi\in S_p^{\dot\sim}}\,\,\,\sum_{\pi'\in[\pi]}\,,
\end{equation}
where $[\cdot]$ denotes an equivalence class and $A^\sim\,(S_p^{\simtwo})$ denotes a
complete set of representatives with respect to the equivalence relation $\sim$ ($\simtwo$).\footnote{Our concepts of equivalence and equivalence class are closely related, but not identical, to these concepts in group theory. We simply use them to split the sums as in \eqref{eq:aux_range_splitting}. As in group theory, our classes are disjoint and cover the entire range of the sums.}

We now insert \eqref{eq:aux_range_splitting} into \eqref{eq:gauge_int}
and use \eqref{eq:property_grass_perm_weing} in the sums over $\alpha'$, $\pi'$, and $\sigma'$. Taking into account the effect of the Grassmann variables in \eqref{eq:part_func1}, we can replace
\begin{equation}
  \label{eq:Igsimple}
  \mathcal{I}^{g\bar g} \to 
\sum_{\alpha\in A^{\sim}}
  \,\,\,
  \sum_{\pi,\sigma\in S_p^{\simtwo}}
  \mathcal{L}_{\alpha\pi\sigma}^{g\bar g}
  I_{\alpha\pi\sigma}^{\text{b}}
  I_{\alpha\pi\sigma}^{\text{e}}
  \,,
\end{equation}
which becomes an equality after contraction with all Grassmann variables connected to the link in question. The link weight is defined as
\begin{equation}
  \mathcal{L}_{\alpha\pi\sigma}^{g\bar g}
  = \left[\prod_{s=1}^{\Nc-1}\frac{s!}{(s+q)!}\right]  |[\alpha]|\sum_{\pi' \in[\pi]}\sum_{\sigma' \in[\sigma]}\, 
  \sgn(\pi'^{-1}\cdot\pi) \,
  \sgn(\sigma'^{-1}\cdot\sigma)
  \Wg_{\Nc}^{q,p}(\pi'\cdot\sigma'^{-1})\,.
\end{equation}
Here, the two sign factors are generated by the commutation of Grassmann variables when transforming $\pi'$ and $\sigma'$ to their representatives $\pi$ and $\sigma$, respectively. The permutations needed to transform the groupings $\alpha'$ to their representatives $\alpha$ always come in pairs (from a $\psi$ at one end and a $\bar\psi$ at the other end of the link), and hence these permutations do not generate sign factors but merely produce a factor of $|[\alpha]|$, which is the cardinality of the equivalence class $[\alpha]$.

What have we gained? The quantity $\mathcal{L}_{\alpha\pi\sigma}^{g\bar g}$ is a number that is independent of the color indices. It can easily be precomputed for all desired combinations of $g$, $\bar g$, $\alpha$, $\pi$, and $\sigma$. The remaining sums over $\alpha,\pi,\sigma$ in \eqref{eq:Igsimple} are now only over the representatives and thus contain much fewer terms than the sums in \eqref{eq:gauge_int}. In the following we will use the shorthand notation
\begin{equation}
  \label{eq:r}
  r \equiv (\alpha,\pi,\sigma)
  \qquad\text{and}\qquad
  \sum_{r} \equiv\sum_{\alpha\in A^{\sim}}
  \,\,\,
  \sum_{\pi,\sigma\in S_p^{\simtwo}}.
\end{equation}

\section{Grassmann integration}\label{sec:Grass_int}

We now consider the nearest-neighbor Grassmann hopping terms in $\mathcal{G}$, see \eqref{eq:part_func1} and \eqref{eq:Grassmann_block}, and turn them into purely local contributions, suitable for a tensor network, based on ideas developed for GHOTRG \cite{Shimizu:2014uva,Bloch:2022vqz}. To this end we introduce a new auxiliary Grassmann variable $\aux_{x,\mu}$ on each link,\footnote{The boundary conditions for the $\aux_{x,\mu}$ are antiperiodic in time and periodic otherwise.} with corresponding differential $d\aux_{x,\mu}$, and integrate out the original Grassmann variables. Since these steps are quite technical we perform the corresponding calculations in \cref{app:Grassm_factor}, where we also Taylor-expand the term $\exp(\SM)$ in the partition function.  We obtain
\begin{equation}\label{eq:Grassmann_evaluation}
  \int \left[\prod_x d\psi_x d\bar\psi_x\right] \mathcal{G}
  = \pif \prod_x K_x\sigma_x\prod_f\frac{(2m_f)^{w_x^f}}{w_x^f!} E_x^f \Theta(w_x^f) \, \delta_{\hin_x,\hout_x}  \,.
\end{equation}
Here, the integral on the right-hand side is over all auxiliary Grassmann variables,
\begin{equation}\label{eq:pif}
  \pif\equiv\prod_{x,\mu}\left(\int\right)^{\F_{x,\mu}}\,,
\end{equation}
where the differentials are part of
\begin{equation}\label{eq:Gxfx}
  K_x \equiv
   \prod_\mu\left(\aux_{x,\mu}\right)^{\F_{x,\mu}}
   \coprod_\mu\left(d\aux_{x,-\mu}\right)^{\F_{x,-\mu}}
\end{equation}
with Grassmann parities\footnote{In earlier papers this Grassmann parity is denoted by $f$ instead of $\F$, but we have already used $f$ as a flavor index.} 
\begin{equation}\label{eq:f}
  \F_{x,\mu}\equiv
  \begin{cases}
    0 & \text{if } k_{x,\mu}+\bar k_{x,\mu} \text{ even}, \\
    1 & \text{if } k_{x,\mu}+\bar k_{x,\mu} \text{ odd}.
  \end{cases}
\end{equation}
Furthermore,
$\sigma_x$ is a local sign factor resulting from permutations of Grassmann variables, given in \eqref{eq:sigma}, \eqref{eq:sigma1} and \eqref{eq:sigma2}. The quantities
\begin{equation}\label{eq:hin_hout}
  \hin_x \equiv \sum_\mu (k_{x,-\mu}^f+\bar k_{x,\mu}^f)\quad\text{and}\quad
  \hout_x \equiv \sum_\mu (\bar k_{x,-\mu}^f+k_{x,\mu}^f)
\end{equation}
are the numbers of incoming and outgoing fermion hopping terms for flavor $f$ at site $x$.
The power $w_x^f$ of $m_f$ in \eqref{eq:Grassmann_evaluation} is called mass occupation number and given by
\begin{equation}\label{eq:wx}
  w_x^f= \Nc - \hin_x  \,.
\end{equation}
The Heaviside function and the Kronecker delta in \eqref{eq:Grassmann_evaluation} lead to the constraint $\hin_x=\hout_x\le \Nc$.
For given site and flavor, the integration of the original Grassmann variables produces the factor
\begin{align}\label{eq:Grassmann_epsilon_site}
  E_x^f
  &=
    \epsilon_{v_x^{f,1},\dots,v_x^{f,w_x^f},
    i_{x,1}^{\kappa_{x,1}^f+1},\dots,i_{x,1}^{\kappa_{x,1}^f+k_{x,1}^f}, \dots,
    i_{x,d}^{\kappa_{x,d}^f+1},\dots,i_{x,d}^{\kappa_{x,d}^f+k_{x,d}^f},
    \ell_{x,-1}^{\bar\kappa_{x,-1}^f+1},\dots,\ell_{x,-1}^{\bar\kappa_{x,-1}^f+\bar k_{x,-1}^f}, \dots,
    \ell_{x,-d}^{\bar\kappa_{x,-d}^f+1},\dots,\ell_{x,-d}^{\bar\kappa_{x,-d}^f+\bar k_{x,-d}^f}}\notag\\
  &\quad\times  \epsilon_{v_x^{f,1},\dots,v_x^{f,w_x^f},
    m_{x,1}^{\bar\kappa_{x,1}^f+1},\dots,m_{x,1}^{\bar\kappa_{x,1}^f+\bar k_{x,1}^f}, \dots,
    m_{x,d}^{\bar\kappa_{x,d}^f+1},\dots,m_{x,d}^{\bar\kappa_{x,d}^f+\bar k_{x,d}^f},
    j_{x,-1}^{\kappa_{x,-1}^f+1},\dots,j_{x,-1}^{\kappa_{x,-1}^f+k_{x,-1}^f}, \dots,
    j_{x,-d}^{\kappa_{x,-d}^f+1},\dots,j_{x,-d}^{\kappa_{x,-d}^f+k_{x,-d}^f}}\,,
\end{align}
where all color indices are associated with site $x$ and flavor $f$. Inside $E_x^f$, we sum over all color indices $v_x^{f,a}$.

Two remarks are in order.
First, the Levi-Civita tensors in $E_x^f$ appear to have many indices. However, because of \eqref{eq:wx} and the subsequent constraint, both Levi-Civita tensors have exactly $\Nc$ indices each. If $k_{x,\pm\mu}^f$ or $\bar k_{x,\pm\mu}^f$ or $w_x^f$ is zero, the corresponding color indices are not present.
Second, the constraint $\hin_x=\hout_x$ can be used to show that $\sum_\mu (\F_{x,\mu}+\F_{x,-\mu})$ is even. This implies that $K_x$, although it contains Grassmann variables, is Grassmann even and thus commuting.

\section{Construction of the tensor network}
\label{sec:tensor_network}

\subsection{Elimination of color degrees of freedom}
\label{sec:elimination_color_dof}

We now consider the full partition function \eqref{eq:part_func1} and insert \eqref{eq:Igsimple} and \eqref{eq:Grassmann_evaluation} to obtain
\begin{equation}
  \label{ZafterGrassmann}
  \ZQCD = \sum_{\bm{k}, \bm{n}} \sum_{\bm{r}} \,
  \mathcal{W}
  \Deltatot
  \pif \prod_x K_x\sigma_x
  \left[\prod_f\frac{(2m_f)^{w_x^f}}{w_x^f!} E_x^f \Theta(w_x^f) \, \delta_{\hin_x,\hout_x}\right]
  \prod_{\mu}
  I_{x,\mu}^{\text{b}}
  I_{x,\mu}^{\text{e}}
  \mathcal{L}_{x,\mu}\,,
\end{equation}
where we defined
\begin{equation}
  I_{x,\mu}^{\text{b}}\equiv (I_{r_{x,\mu}}^\text{b})^{\bm{i}_{x,\mu}\bm{m}_{x,\mu}}\,,\quad
  I_{x,\mu}^{\text{e}}\equiv (I_{r_{x,\mu}}^\text{e})^{\bm{j}_{x,\mu}\bm{\ell}_{x,\mu}}\,,\quad
  \mathcal{L}_{x,\mu}\equiv\mathcal{L}_{r_{x,\mu}}^{g_{x,\mu}\bar g_{x,\mu}}
\end{equation}
for simplicity and introduced the field $\bm{r}\equiv(r_{x,\mu}| \, \forall \, x,\mu)$, again with periodic boundary conditions.

Let us briefly discuss this intermediate result. First, for every link, the gauge integral results in a sum over $r$ (i.e., over representatives $\alpha$, $\pi$, and $\sigma$, see \eqref{eq:r}) for this link.
The range of $r$ for a given link depends on $k$, $\bar k$, and $\vec{n}$ for this link, where $\vec{n}$ contains all occupation numbers of plaquettes which include that link. If the $\SU(\Nc)$ condition $|g-\bar g|/\Nc\in\mathbb{N}_0$ is not satisfied, the range of $r$ is zero and the tuple $(k,\bar k, \vec{n})$ does not contribute to the partition function. Second, the integral over the original Grassmann variables has been replaced by an integral over the auxiliary Grassmann variables, which are color- and flavorless.  The original Grassmann variables were coupled to neighbors by hopping terms. In contrast, the auxiliary Grassmann variables, which live on the links, are decoupled. This is the main achievement of \cref{sec:Grass_int}.

A given term in the sum over $\bm k$, $\bm n$, and $\bm r$ in \eqref{ZafterGrassmann} contains factors from all sites, links, and plaquettes of the lattice. To construct a network of local tensors we need to gather the factors corresponding to a single site and contract their color indices. This can be done for each lattice site separately, as we show in the following.

We first rewrite $\Deltatot$ in \eqref{Delta} in terms of local objects.
To this end, we shift the first factor in \eqref{eq:Delta_a} from site $x$ to site $x-\hat\mu$, the second factor from site $x$ to site $x-\hat\mu-\hat\nu$, and the third factor from site $x$ to site $x-\hat\nu$, respectively, to obtain\footnote{Note that each Kronecker delta just enforces the equality of any two color indices at a plaquette corner.}
\begin{align}\label{eq:plaq_delta_site}
\Deltatot=\prod_x\Delta_x\quad\text{with}\quad\Delta_x &\equiv \prod_{\mu\neq\nu}
	\left(\prod_{a=1}^{n_{x-\hat\mu,\mu\nu}} \delta_{j_{x,-\mu}^{a+\of{+\nu}_{x,-\mu}},i_{x,\nu}^{a+\of{-\mu}_{x,\nu}}}\right)
	\left(\prod_{a=1}^{n_{x-\hat\mu-\hat\nu,\mu\nu}} \delta_{j_{x,-\nu}^{a+\of{-\mu}_{x,-\nu}},\ell_{x,-\mu}^{a+\bof{-\nu}_{x,-\mu}}}\right)
	\notag\\
	&\qquad\times
	\left(\prod_{a=1}^{n_{x-\hat\nu,\mu\nu}} \delta_{m_{x,\mu}^{a+\bof{-\nu}_{x,\mu}},\ell_{x,-\nu}^{a+\bof{+\mu}_{x,-\nu}}}\right)
	\left(\prod_{a=1}^{n_{x,\mu\nu}} \delta_{m_{x,\nu}^{a+\bof{+\mu}_{x,\nu}},i_{x,\mu}^{a+\of{+\nu}_{x,\mu}}}\right)
	.
\end{align}
This expression depends on all color indices associated with site $x$ that originate from the plaquette action.
Analogously, the contributions depending on the color indices in the gauge-link integrals in \eqref{ZafterGrassmann} can be rewritten locally by shifting the sites in $I_{x,\mu}^{\text{e}}$ from $x$ to $x-\hat\mu$,
\begin{equation}
	I_x \equiv
	\prod_{\mu}
	I_{x,\mu}^{\text{b}}
	I_{x,-\mu}^{\text{e}}\,,
\end{equation}
so that
\begin{equation}
	\prod_x
	\prod_{\mu}
	I_{x,\mu}^{\text{b}}
	I_{x,\mu}^{\text{e}}
	=\prod_x I_x
	\,.
\end{equation}
The local expression $I_x$ depends on all color indices (originating from hopping terms and the plaquette action) that are associated with site $x$.\footnote{A similar decoupling is obtained in Ref.~\cite{Gagliardi:2019cpa} using so-called decoupling operators.} We now define the local color factor
\begin{equation}\label{eq:Cx}
  C_x=
  \Delta_x
  I_x
  \prod_f E_x^f
  \,,
\end{equation}
which only depends on the variables associated with site $x$,
\begin{subequations}\label{eq:occ_x}
  \begin{align}
    \bm{k}_x  &= (k_{x,-\mu}^f,k_{x,\mu}^f, \bar k_{x,-\mu}^f,\bar k_{x,\mu}^f \,|\, \forall \mu,f) \,,
    \\
    \bm{n}_x &= (n_{x,\mu\nu},n_{x-\hat\mu,\mu\nu},n_{x-\hat\nu,\mu\nu},n_{x-\hat\mu-\hat\nu,\mu\nu} \,|\, \forall \mu\neq\nu)\,,
    \\
    \bm{r}_x &=(r_{x,-\mu},r_{x,\mu} \,|\, \forall \mu)\,.
  \end{align}
\end{subequations}
Every color index corresponding to site $x$ appears exactly twice on the right-hand side of \eqref{eq:Cx}.
We computed $C_x$ analytically by explicitly summing over the color indices using \texttt{sympy} \cite{10.7717/peerj-cs.103}.
The partition function \eqref{ZafterGrassmann} then becomes
\begin{equation}
  \label{ZafterColor}
  \ZQCD = \sum_{\bm{k}, \bm{n}} \sum_{\bm{r}}
  \mathcal{W}
  \pif
  \prod_x   K_x \sigma_x  C_x
\left[\prod_f \frac{(2m_f)^{w_x^f}}{w_x^f!} \Theta(w_x^f) \delta_{\hin_x,\hout_x} \right]
  \prod_\mu \mathcal{L}_{x,\mu}\,.
\end{equation}

\subsection{Tensor formulation}
\label{sec:tensor}

In the partition function we have several summation variables, where $k_{x,\mu}^f$ and $\bar k_{x,\mu}^f$ are link occupation numbers, $n_{x,\mu\nu}$ is a plaquette occupation number, and $r_{x,\mu}$ is a link variable whose range depends on $k$, $\bar k$, and $n$. We rewrite the partition function \eqref{ZafterColor} as a tensor network
\begin{equation}\label{Zplaqtens}
  \ZQCD = \sum_{\bm{k}, \bm{n}} \sum_{\bm{r}}\pif \prod_x \overline{T}_x K_x
\end{equation}
with the local numerical tensor
\begin{equation}\label{eq:Tbarx}
  \overline{T}_x
  \equiv  \sigma_x C_x  W_x \left[\prod_f \frac{(2m_f)^{w_x^f}}{w_x^f!}\Theta(w_x^f) \, \delta_{\hin_x,\hout_x} \right]
     \left[\prod_\mu\text{sgn}(\mathcal{L}_{x,\mu})\right]
    \prod_{\mu=\pm1}^{\pm d}\sqrt{|\mathcal{L}_{x,\mu}|}
\end{equation}
whose tensor indices are $\bm k_x$, $\bm n_x$, and $\bm r_x$.
To obtain this tensor-network structure we also factorized the weight $\mathcal{W}$ of \eqref{numpref} into local contributions by writing
$\mathcal{W} = \prod_x W_x$
with
\begin{align}\label{eq:Wx}
  W_x = 
        \left[\prod_{\mu\neq\nu} \frac{{\left(\frac{\beta}{2\Nc}\right)}^{n_{x,\mu\nu}+n_{x-\hat\mu,\mu\nu}+n_{x-\hat\mu-\hat\nu,\mu\nu}+n_{x-\hat\nu,\mu\nu}}}{n_{x,\mu\nu}! n_{x-\hat\mu,\mu\nu}! n_{x-\hat\mu-\hat\nu,\mu\nu}! n_{x-\hat\nu,\mu\nu}!}\right]^{\frac14}
  \left[\prod_{\mu}\eta_{x,\mu}^{\F_{x,\mu}}\right]
\left[\prod_{\mu=\pm1}^{\pm d}
\prod_f\sqrt{\frac{e^{\mu_f(k_{x,\mu}^f-\bar k_{x,\mu}^f)\delta_{|\mu|,1}}}{k_{x,\mu}^f!\bar k_{x,\mu}^f!}}\right]
  ,
\end{align}
where all numerical quantities on a link were evenly distributed via square roots to both endpoints, except for the staggered phases and the sign of $\mathcal{L}_{x,\mu}$ to avoid complex entries, and all numerical quantities on a plaquette were distributed via fourth roots to its four corners.
Similarly, the link weight $|\mathcal{L}_{x,\mu}|$ was distributed via square roots to the two tensors at the endpoints of the link.
Recall that $\mu_f$ in the last factor of \eqref{eq:Wx} is the chemical potential for flavor $f$.

\begin{figure}
  \centering
  \includegraphics{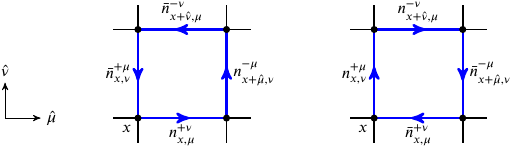}
  \caption{Introducing edge variables for plaquettes $\UP_{x,\mu\nu}$ (left) and $\UP_{x,\nu\mu}$ (right).}
  \label{fig:plaq_to_link}
\end{figure}

In standard tensor-renormalization-group methods, two tensors at neighboring sites are contracted by summing over the shared link index on the link connecting the two sites. However, 
the sum in \eqref{Zplaqtens} involves plaquette occupation numbers $n$, which are shared by four tensors. A summation over plaquette occupation numbers therefore does not correspond to a contraction of two local tensors $T_x$. A similar argument applies to $r$, whose range is determined by $n$.
To be able to still apply tensor-renormalization-group methods, one option would be to develop a more general blocking procedure for a tensor network containing indices that are shared by more than two tensors. While this would indeed be an interesting option to explore in future work, here we follow a more traditional route. We convert the plaquette occupation numbers $n_{x,\mu\nu}$
to link variables and use the GHOTRG method developed for the infinite-coupling case \cite{Bloch:2022vqz}. This is implemented by introducing a link variable for each edge of a plaquette, called ``edge variable'' in the following, and requiring that all four edge variables around a plaquette take on the same value for terms that contribute to the partition function. For the plaquette $\UP_{x,\mu\nu}$ we introduce $n^{+\nu}_{x,\mu}$, $n^{-\mu}_{x+\hat\mu,\nu}$, $\bar n^{-\nu}_{x+\hat\nu,\mu}$, and $\bar n^{+\mu}_{x,\nu}$, as illustrated in \cref{fig:plaq_to_link}. The superscript $\pm\mu$ indicates that the plaquette extends in the positive/negative $\mu$-direction perpendicular to the link under consideration, while $n$ and $\bar n$ correspond to $U$ and $U^\dagger$, respectively.

\begin{figure}
\centering
\includegraphics{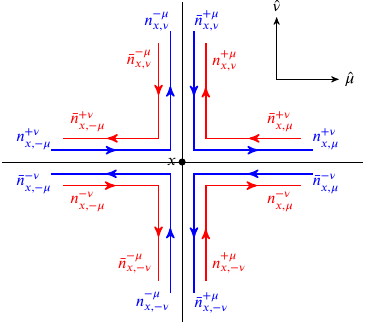}
\caption{Enumeration of the edge variables associated with site $x$.}
\label{fig:edge_indices}
\end{figure}

Formally, we rewrite the sum of a generic function $f$ over a plaquette occupation number as
\allowdisplaybreaks[0]
\begin{align}
  \sum_{n_{x,\mu\nu}} f(n_{x,\mu\nu})
  &= \sum_{n_{x,\mu\nu}} f(n_{x,\mu\nu})
    \sum_{n^{+\nu}_{x,\mu}} \delta_{n^{+\nu}_{x,\mu},n_{x,\mu\nu}} 
    \sum_{n^{-\mu}_{x+\hat\mu,\nu}} \delta_{n^{-\mu}_{x+\hat\mu,\nu},n_{x,\mu\nu}}
    \sum_{\bar n^{-\nu}_{x+\hat\nu,\mu}} \delta_{\bar n^{-\nu}_{x+\hat\nu,\mu},n_{x,\mu\nu}}
    \sum_{\bar n^{+\mu}_{x,\nu}} \delta_{\bar n^{+\mu}_{x,\nu},n_{x,\mu\nu}} \notag\\
  &= \sum_{n^{+\nu}_{x,\mu},n^{-\mu}_{x+\hat\mu,\nu},\bar n^{-\nu}_{x+\hat\nu,\mu},\bar n^{+\mu}_{x,\nu}}
    \delta_{n^{+\nu}_{x,\mu},n^{-\mu}_{x+\hat\mu,\nu}} 
    \delta_{n^{-\mu}_{x+\hat\mu,\nu},\bar n^{-\nu}_{x+\hat\nu,\mu}}
    \delta_{\bar n^{-\nu}_{x+\hat\nu,\mu},\bar n^{+\mu}_{x,\nu}}
    \delta_{\bar n^{+\mu}_{x,\nu},n^{+\nu}_{x,\mu}} f(\text{edge variable})
    \,,
    \label{eq:deltas}
\end{align}
\allowdisplaybreaks[3]
where we have eliminated the original sum over $n_{x,\mu\nu}$ in the second step. Although one of the Kronecker deltas in the second line is redundant, it is useful for the tensor formulation to use this symmetrized form. Effectively, each Kronecker delta imposes the equality of the edge variables around a plaquette corner.
Contributions to the partition function that depend on $n_{x,\mu\nu}$ will only be nonzero if all four edge variables are equal, as they represent the same plaquette occupation number. Therefore, every occurrence of $n_{x,\mu\nu}$ can be replaced by any of the edge variables $n^{+\nu}_{x,\mu}$, $n^{-\mu}_{x+\hat\mu,\nu}$, $\bar n^{-\nu}_{x+\hat\nu,\mu}$, or $\bar n^{+\mu}_{x,\nu}$.
We gather all these single-plaquette constraints on the lattice in
\begin{equation}
  \DeltaP \equiv
  \prod_{x,\mu\neq\nu} 
  \delta_{n^{+\nu}_{x,\mu},n^{-\mu}_{x+\hat\mu,\nu}}
  \delta_{n^{-\mu}_{x+\hat\mu,\nu},\bar n^{-\nu}_{x+\hat\nu,\mu}}
  \delta_{\bar n^{-\nu}_{x+\hat\nu,\mu},\bar n^{+\mu}_{x,\nu}}
  \delta_{\bar n^{+\mu}_{x,\nu},n^{+\nu}_{x,\mu}} \,.
  \label{eq:plaq_occ_delta}
\end{equation}
This can be rewritten in terms of local objects by separately shifting the site in each Kronecker delta, which results in
\begin{equation}
  \label{DeltaP}
  \DeltaP = \prod_{x}\DeltaP_x
  \quad\text{with}\quad
  \DeltaP_x = \prod_{\mu\neq\nu}
  \delta_{n^{+\nu}_{x,-\mu},n^{-\mu}_{x,\nu}}
  \delta_{n^{-\mu}_{x,-\nu},\bar n^{-\nu}_{x,-\mu}}
  \delta_{\bar n^{-\nu}_{x,\mu},\bar n^{+\mu}_{x,-\nu}}
  \delta_{\bar n^{+\mu}_{x,\nu},n^{+\nu}_{x,\mu}}\,, 
\end{equation}
where $\DeltaP_x$ depends on all edge variables associated with site $x$. We thus have four new variables $n^{+\nu}_{x,\mu}$, $\bar n^{+\nu}_{x,\mu}$, $\bar n^{-\nu}_{x,\mu}$, $n^{-\nu}_{x,\mu}$ on every link $(x,\mu)$ of the lattice for each plane $\mu\nu$, see \cref{fig:edge_indices}. The range of $r_{x,\mu}$ depends on the plaquette occupation numbers through $g_{x,\mu}$ and $\bar g_{x,\mu}$ defined in \eqref{eq:g}. Because of the Kronecker deltas in \eqref{eq:deltas} the plaquette occupation numbers in \eqref{eq:g} can be replaced by edge variables defined on the link $(x,\mu)$. Thus $r_{x,\mu}$ now only depends on link variables.

Now that all variables are defined on the links we can combine the various link variables in a compound link index $j_{x,\mu}=((k_{x,\mu}^f,\bar k_{x,\mu}^f\,|\,\forall f),(n^{\pm\nu}_{x,\mu},\bar n^{\pm\nu}_{x,\mu} \,|\,\forall \nu\neq\mu),r_{x,\mu})$
and rewrite the partition function \eqref{Zplaqtens} as
\begin{equation}\label{eq:final_tensor_network}
  \ZQCD = \sum_{\bm{j}}\pif \prod_x T_x K_x \,,
\end{equation}
where $\bm{j}\equiv(j_{x,\mu}| \,\forall\, x, \mu)$.
The numerical tensor is given by
\begin{equation}\label{tensor-def}
  T_x\equiv  \DeltaP_x \overline T_x(\bm n_x\rightarrow \text{edge variables})
\end{equation}
and depends on $j_{x,\pm\mu}$ $(\mu=1,\ldots,d)$. Recall that $\overline T_x$, defined in \eqref{eq:Tbarx}, depends on the plaquette occupation numbers in $\bm n_x$, see \eqref{eq:occ_x}. We have a choice of how to replace them by edge variables defined on links connected to $x$, as described after \eqref{eq:deltas}, but the result does not depend on what choice we make.

\subsection{Restriction of the edge variables}
\label{sec:trunc}

To contract the complete tensor network and compute the partition function and thermodynamical observables using the GHOTRG \cite{Bloch:2022vqz} (or another) method, the bond dimension of the initial tensor has to be finite. While $k$ and $\bar k$ are limited by the number of colors $\Nc$, and $r$ is finite as well for a given gauge integral, the plaquette occupation numbers $n$ run from zero to infinity in the original expansion. In the tensor formulation, the plaquette occupation numbers have been turned into edge variables, which also run from zero to infinity.

To make the range of the tensor indices $j_{x,\mu}$ finite, we need to restrict the edge variables. More precisely, we introduce a maximal order $\nmax$ and remove, for a given link $(x,\mu)$, all combinations of edge variables that do not satisfy the criterion
\begin{equation}\label{eq:link_crit}
  \sum_{{\nu=\pm1}\atop{|\nu|\ne\mu}}^{\pm d}(n_{x,\mu}^\nu+\bar n_{x,\mu}^\nu)\leq \nmax\,.
\end{equation}
With this restriction, all configurations of order $\beta^n$ with $0 \leq n \leq \nmax$ are included in the calculation, while higher orders will be incomplete.
Note that we can substantially speed up the tensor construction by setting to zero the entries of the initial tensor for which\footnote{The factor of $1/2$ arises since two edge variables correspond to the same plaquette occupation number, see \cref{fig:edge_indices}.}
\begin{equation}\label{eq:site_crit}
  \frac12\sum_{\mu=\pm1}^{\pm d} \sum_{{\nu=\pm1}\atop{|\nu|\ne|\mu|}}^{\pm d}(n_{x,\mu}^\nu+\bar n_{x,\mu}^\nu) > \nmax \,,
\end{equation}
which correspond to tensor entries, and hence configurations, of order $\beta^n$ with $n>\nmax$. Although this step does not reduce the size of the initial tensor, it is essential to keep the runtimes under control.\footnote{After choosing $\nmax$, the bond dimension of the initial tensor is fixed. However, for an efficient numerical computation, especially for higher orders in $\beta$, the initial tensor may additionally be truncated to a particular bond dimension $D$, using an HOSVD-inspired truncation procedure analogous to that of the GHOTRG method.} Note that the conditions \eqref{eq:link_crit} and \eqref{eq:site_crit} need to be slightly altered when the lattice consists of a single site in any direction to avoid double counting when taking the boundary conditions into account.

\section{Strategies to expand observables and results for small lattices}
\label{sec:prelim}

In the following we present first tensor-network results for the strong-coupling expansion of two-dimensional QCD with three colors and one flavor, i.e., $\Nc=3$ and $\Nf=1$. We omit the flavor index in the following.

As we eventually want to use the initial tensor, constructed in this paper, as input to the GHOTRG method on large lattices, it is enlightening to first investigate the exact results that can be obtained with this tensor on a $2\times2$ lattice.
Starting from the initial tensor $T_xK_x$ in \eqref{eq:final_tensor_network} and using the tensor  \eqref{tensor-def} to order $\nmax$ by imposing the constraints \eqref{eq:link_crit} and \eqref{eq:site_crit}, the partition function on a $2\times2$ lattice can be computed exactly to order $\nmax$,
by exhaustively enumerating all terms in the partition function \eqref{eq:final_tensor_network} in an efficient way. Each such term corresponds to a specific order $n$ in $\beta$, and all nonzero contributions satisfy $n\leq\nmax$.\footnote{On a $2\times2$ lattice no terms with order $n>\nmax$ appear in the exact computation of $Z$. Indeed, for this lattice size the edge occupation numbers of the four tensors are shared in such a way that if the initial tensors have maximal order $\nmax$, \eqref{eq:site_crit} ensures that all nonzero contributions to the fully contracted tensor network have at most order $\nmax$ as well. On larger lattices higher order contributions with $n>\nmax$ also contribute to $Z$.}
After gathering the terms order by order we can compute all the coefficients $Z_n$ separately and write the partition function as
\begin{align}\label{eq:Zexp}
  Z^{(\nmax)} = \sum_{n=0}^{\nmax} Z_n \beta^n \,.
\end{align}
Analytical expressions for $Z_n$ as functions of $m$ and $\mu$ are given in \cref{app:Zn} up to $n=4$.

To compare tensor results with those of Monte Carlo simulations we will compute the chiral condensate,
\begin{equation}
  \label{eq:condensate}
  \Sigma \equiv 2\braket{\bar\psi\psi}=\frac1V\frac{\partial \ln Z}{\partial m}  \,,
\end{equation}
where the factor of $2$ is due to the rescaling of the Grassmann variables mentioned in footnote~\ref{footnote}.
A natural way to perform a tensor-network computation of the chiral condensate is to rewrite \eqref{eq:condensate} as
\begin{equation}
  \label{eq:cc_A}
  \Sigma_A^{(\nmax)}
  =\frac1V\frac{\partial \ln Z^{(\nmax)}}{\partial m}
  = \frac1{VZ^{(\nmax)}}  \frac{\partial Z^{(\nmax)}}{\partial m} \,,
\end{equation}
where we approximated $Z$ by the polynomial $Z^{(\nmax)}$.
The derivative can be computed analytically for the $2\times2$ lattice as the coefficients $Z_n$  are known as explicit functions of $m$ and $\mu$, and \eqref{eq:cc_A} becomes a ratio of polynomials in $\beta$.

Alternatively, we can expand \eqref{eq:cc_A} or, equivalently, $\ln Z$ in $\beta$ and truncate that expansion to order $\nmax$ when computing the observable.  We therefore expand the free energy density $f=-\ln Z/V$ in the form (disregarding the sign)
\begin{equation}\label{eq:logZexp}
	\frac{\ln Z(\beta)}{V}=\sum_{n=0}^{\nmax} f_n \beta^{n}+\mathcal{O}(\beta^{\nmax+1})\,.
\end{equation}
For example, up to order three we obtain with $\Zbar_n\equiv Z_n/Z_0$
\begin{equation}
	\frac{\ln Z(\beta)}{V} = \frac{\ln Z_0}{V}
	+\frac{\beta}{V}\Zbar_1+\frac{\beta^2}{V}\left(\Zbar_2-\frac12\Zbar_1^2\right)
	+\frac{\beta^3}{V}\left(\Zbar_3-\Zbar_1\Zbar_2+\frac{1}{3}\Zbar_1^3\right)
	+\mathcal{O}(\beta^{4})\,.
	\label{logZexp}
\end{equation}
Taking the derivative with respect to $m$ yields
\begin{equation}
	\Sigma_B^{(\nmax)} = \sum_{n=0}^{\nmax} \Sigma_n\beta^n\quad\text{with}\quad \Sigma_n\equiv\frac{\partial f_n}{\partial m} \,.
	\label{eq:cc_expansion}
\end{equation}
In the following we will refer to the alternatives \eqref{eq:cc_A} and \eqref{eq:cc_expansion} as ``expansion $A$'' and ``expansion $B$'', respectively.

\begin{figure}
\centering
\includegraphics[height=55mm]{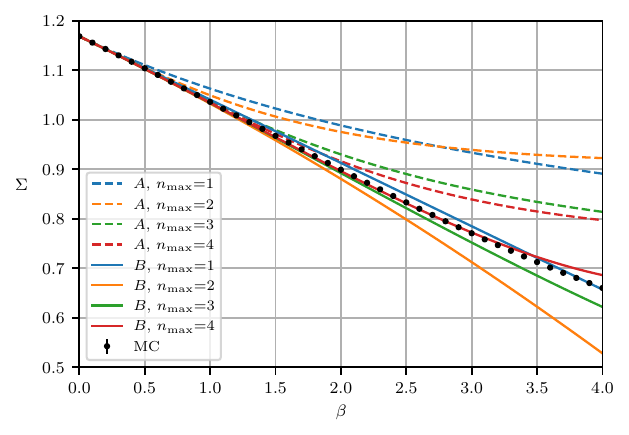}\hfill
\includegraphics[height=55mm]{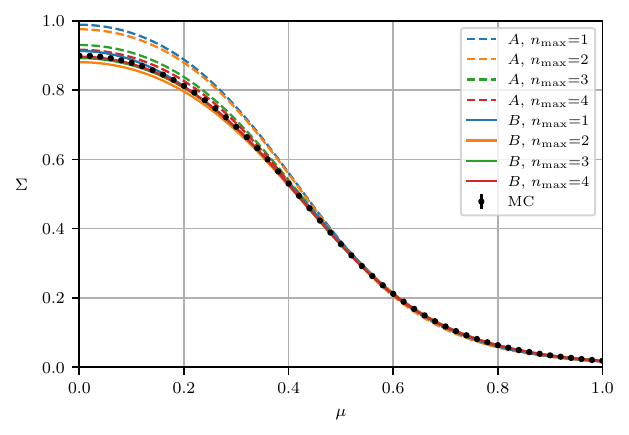}
\caption{Chiral condensate as a function of $\beta$ (left, for $\mu=0$) and $\mu$ (right, for $\beta=2$) for $m=0.1$ on a $2\times2$ lattice. We compare Monte Carlo (MC) results with the two tensor-network expansions $\Sigma_A^{(\nmax)}$ and $\Sigma_B^{(\nmax)}$ of \eqref{eq:cc_A} and \eqref{eq:cc_expansion}, respectively.\label{fig:cc_2x2} The MC results for $\mu\ne0$ were obtained through phase-quenched reweighting, with an average phase factor between 0.81 and 1.}
\end{figure}

In \cref{fig:cc_2x2} we compare both expansions for the chiral condensate, with $\nmax=1,2,3,4$, to the full Monte Carlo results for $m=0.1$.
In the left plot, in which $\mu=0$, we observe that expansion $A$ only agrees with the Monte Carlo results for small values of $\beta$ and deviates quite substantially as $\beta$ increases.\footnote{These results are in agreement with those of Ref.~\cite{Gagliardi:2019cpa}, where $\Sigma/2$ is plotted.}
For expansion $B$, we observe a better agreement with the Monte Carlo results over a larger range in $\beta$.
In the right plot we see that expansion $B$ outperforms expansion $A$ also for $\mu\ne0$.
It thus seems preferable to expand the observable itself in $\beta$ and truncate it to the same order $\nmax$ as the partition function in \eqref{eq:Zexp}.

\begin{figure}
\centering
\includegraphics[height=55mm]{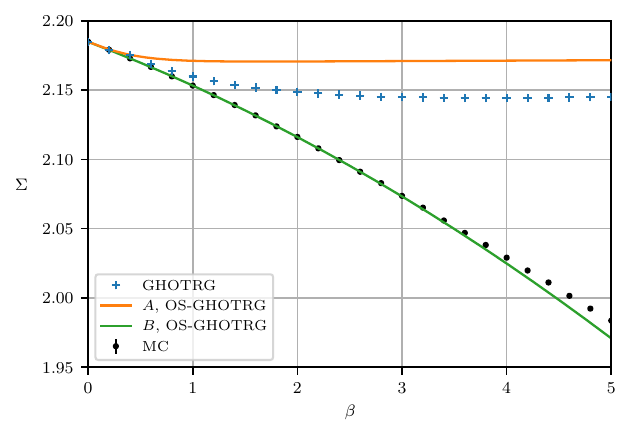}
\caption{\label{fig:os-ghotrg}Chiral condensate versus $\beta$ for $L=8$, $m=0.5$, $\mu=0$ using the standard GHOTRG method with initial tensor truncated to $\nmax=2$ and the OS-GHOTRG method with expansions $A$ and $B$ up to order $\nmax=2$, compared to Monte Carlo data. For expansion $A$ we observe the plateau for large $\beta V$ derived earlier. As expected, expansion $B$ agrees well with the Monte Carlo data over a larger range in $\beta$.}
\end{figure}

Let us compare expansions $A$ and $B$ on theoretical grounds for arbitrary volume.
For the free-energy density to be finite in the infinite-volume limit, the coefficients $f_n$ in \eqref{eq:logZexp} must remain finite as $V\to\infty$. Looking at \eqref{logZexp} term by term, this implies that the ratios $\Zbar_n$ must behave like $V^n$ for large $V$, and that cancellations must take place in every order $n\ge2$ to ensure that the $f_n$ are finite for large $V$. 
Let us now consider expansion $A$, which uses $\ln Z^{(\nmax)}$ in \eqref{eq:cc_A}. If we were to expand $\ln Z^{(\nmax)}$ in $\beta$ (which is not actually done in expansion $A$) we would also obtain terms of order $\beta^n$ with $n>\nmax$, but these higher-order contributions would be incomplete and would have the wrong volume dependence because the cancellations just discussed cannot take place. Moreover, because $\Zbar_n\propto V^n$ for large $V$, \eqref{eq:Zexp} with $Z_0$ factored out is then effectively an expansion in powers of $\beta V$.
For large $\beta V$,
$Z^{(\nmax)}$ is dominated by the term with $n=\nmax$. In this case, \eqref{eq:cc_A} with $Z^{(\nmax)}\approx Z_0\Zbar_\nmax\beta^\nmax$ results in  $\Sigma_A^{(\nmax)}\approx\Sigma_0+\frac1V\partial_m\ln\Zbar_\nmax$, which is independent of $\beta$.
In summary, we expect expansion $B$ to perform better than expansion $A$ when compared to Monte Carlo results for large $\beta V$ (and $\beta$ not too large),
which is consistent with our results for $L=2$ (see \cref{fig:cc_2x2}) and $L=8$ (see \cref{fig:os-ghotrg}). 

In order to use \eqref{eq:cc_expansion} for larger lattices we need to be able to compute the coefficients $Z_n$ of the partition function obtained from the tensor network for such lattices.
However, when applying the standard GHOTRG method to contract the tensor network numerically, we obtain $Z$ (and thus also $\ln Z$ and the thermodynamical observables) as a numerical function of $\beta$, and not as an explicit expansion in $\beta$.
We tried to determine the expansion coefficients $Z_n$ by fitting tensor data to polynomials in $\beta$, but this attempt did not work well on larger lattices since the tensor data have limited precision for realistic bond dimensions, and since on larger lattices the completely contracted tensor network will contain many incomplete higher-order contributions ($n>\nmax$) even though the initial tensor is truncated to order $\nmax$.\footnote{Similar to expansion $A$, the incomplete higher-order terms in the standard GHOTRG method also have a wrong volume dependence, and $\Sigma$ tends to a plateau, see \cref{fig:os-ghotrg}.}

To be able to directly compute the expansion coefficients $Z_n$ we have developed a modified version of GHOTRG, which we call order-separated GHOTRG (OS-GHOTRG). Once the coefficients $Z_n$ are known, we can compute expansion coefficients for all thermodynamical observables. 
In \cref{fig:os-ghotrg} we show that the results for $L=8$ obtained with expansion $B$ using the OS-GHOTRG method agree much better with Monte Carlo data than those from the standard GHOTRG method and from expansion $A$ using the OS-GHOTRG method (the latter two are identical for $L=2$).
The OS-GHOTRG method is quite complex and will be detailed in a forthcoming paper.

\section{Conclusions and outlook}
\label{sec:conclusions}

We derived a tensor-network formulation to compute the partition function of QCD in the strong-coupling expansion for arbitrary space-time dimension, number of colors, and number of staggered fermion flavors.
We first performed Taylor expansions of the exponentials involving the gauge and fermion actions, which generate fermion and plaquette occupation numbers.
We then integrated out the gauge field and observed that, due to the presence of Grassmann variables at both ends of every link, the subsequent calculation can be simplified, i.e., the sum over permutations and groupings of color indices in the gauge integral can be reduced significantly.
In the next step we introduced colorless auxiliary Grassmann variables on the links of the lattice, which enabled us to integrate out the original Grassmann variables.
Finally, we summed over all color indices to obtain a partition function given by the full contraction of a tensor network of mixed numerical and Grassmann tensors.
For a practical computation of the partition function using the tensor network we truncated the initial tensor to a maximal order $\nmax$ in $\beta^n$.

We presented results from an exact analytical computation of the partition function up to order $\nmax=4$ on a $2\times 2$ lattice. These results primarily served to illustrate the use of the initial tensor constructed in this paper. However, we also observed that the results of the tensor-network method strongly depend on how the expansion of $Z$ is used to compute observables.  The analytical results for $L=2$, numerical results for $L=8$, and a theoretical analysis for arbitrary volume show that reliable results in a significant range of $\beta$ require an expansion of $\ln Z$ or the observable itself, and not only an expansion of $Z$. This in turn requires knowledge of the expansion coefficients $Z_n$, also for larger lattices where they are not available through standard tensor-network methods. In the next paper of this series, we will solve this problem by presenting the OS-GHOTRG method, which allows us to compute the $Z_n$ explicitly.

\appendix

\section{Grassmann contribution}
\label{app:Grassm_factor}

In this appendix we integrate out all the original Grassmann variables in $\mathcal{G}$ of \eqref{eq:Grassmann_block}, i.e., we compute
\begin{align}\label{eq:Gint0}
  \int \left[\prod_x d\psi_x d\bar\psi_x\right] \mathcal{G}\,,
\end{align}
to arrive at an expression that is suitable for the tensor-network formulation.
A basic ingredient will be the relation
\begin{equation}
  \prod_{i=1}^{k}\alpha^i\beta^i
  =\prod_{i=1}^k\alpha^i\coprod_{i=1}^k\beta^i
  =\coprod_{i=1}^k\alpha^i\prod_{i=1}^k\beta^i
  \label{reorder}
\end{equation}
which is valid if, for every $i$, the product $\alpha^i\beta^i$ is commuting. The relation is most easily shown by moving suitable pairs of Grassmann variables. The symbol $\coprod$ denotes the reverse product defined in \eqref{eq:coprod}.

The expression for $\mathcal{G}_{x,\mu}$ in \eqref{eq:Grassmann_block} can be manipulated as follows,
\begin{align}\label{eq:Nested_Grass1}
  \mathcal{G}_{x,\mu}
  &=\left[\prod_f\prod_{a=1}^{k_{x,\mu}^f} \bar{\psi}_{x}^{f,i_{x,\mu}^{a+\kappa_{x,\mu}^f}}
    \psi_{x+\hat\mu}^{f,j_{x,\mu}^{a+\kappa_{x,\mu}^f}}\right]
    \left[\prod_f\prod_{a=1}^{\bar k_{x,\mu}^f} \psi_{x}^{f,m_{x,\mu}^{a+\bar\kappa_{x,\mu}^f}}
    \bar{\psi}_{x+\hat\mu}^{f,\ell_{x,\mu}^{a+\bar\kappa_{x,\mu}^f}}\right]\notag\\
  &=\left[\prod_f\coprod_{a=1}^{k_{x,\mu}^f} \bar{\psi}_{x}^{f,i_{x,\mu}^{a+\kappa_{x,\mu}^f}}
    \prod_{a=1}^{k_{x,\mu}^f} \psi_{x+\hat\mu}^{f,j_{x,\mu}^{a+\kappa_{x,\mu}^f}}\right]
    \left[\prod_f\prod_{a=1}^{\bar k_{x,\mu}^f} \psi_{x}^{f,m_{x,\mu}^{a+\bar\kappa_{x,\mu}^f}}
    \coprod_{a=1}^{\bar k_{x,\mu}^f} \bar{\psi}_{x+\hat\mu}^{f,\ell_{x,\mu}^{a+\bar\kappa_{x,\mu}^f}}\right]\notag\\
  &=\left[\coprod_f\coprod_{a=1}^{k_{x,\mu}^f} \bar{\psi}_{x}^{f,i_{x,\mu}^{a+\kappa_{x,\mu}^f}}\right]
    \left[\prod_f\prod_{a=1}^{k_{x,\mu}^f} \psi_{x+\hat\mu}^{f,j_{x,\mu}^{a+\kappa_{x,\mu}^f}}\right]
    \left[\prod_f\prod_{a=1}^{\bar k_{x,\mu}^f} \psi_{x}^{f,m_{x,\mu}^{a+\bar\kappa_{x,\mu}^f}}\right]
    \left[\coprod_f\coprod_{a=1}^{\bar k_{x,\mu}^f} \bar{\psi}_{x+\hat\mu}^{f,\ell_{x,\mu}^{a+\bar\kappa_{x,\mu}^f}}\right]\notag\\
  &=\left[\coprod_f\coprod_{a=1}^{k_{x,\mu}^f} \bar{\psi}_{x}^{f,i_{x,\mu}^{a+\kappa_{x,\mu}^f}}\right]
    \left[\prod_f\prod_{a=1}^{\bar k_{x,\mu}^f} \psi_{x}^{f,m_{x,\mu}^{a+\bar\kappa_{x,\mu}^f}}\right]
    \left[\coprod_f\coprod_{a=1}^{\bar k_{x,\mu}^f} \bar{\psi}_{x+\hat\mu}^{f,\ell_{x,\mu}^{a+\bar\kappa_{x,\mu}^f}}\right]
    \left[\prod_f\prod_{a=1}^{k_{x,\mu}^f} \psi_{x+\hat\mu}^{f,j_{x,\mu}^{a+\kappa_{x,\mu}^f}}\right].
\end{align}
In the first step, we have redistributed the product over flavors in \eqref{eq:Grassmann_block} by reordering the commuting brackets.\footnote{For simplicity, we say ``bracket'' instead of ``bracketed term''.}  In the second step, we applied \eqref{reorder} to the products over $a$.\footnote{We use the reverse product for $\bar\psi$ and the ordinary product for $\psi$ throughout.} In the third step, we applied \eqref{reorder} to the products over $f$, which is justified since both factors in each product are either commuting or anticommuting depending on the value of $k_{x,\mu}^f$ or $\bar k_{x,\mu}^f$.
In the last step, we moved the product of the last two brackets, which is commuting, between the first and second bracket, so that the Grassmann variables are now sorted by site.

As the Grassmann variables in the last two brackets live on a different site, we need to reorder the brackets in the product over $(x,\mu)$ in \eqref{eq:Grassmann_block} in order to gather all Grassmann variables living on a single site. However, the brackets are only commuting when $k_{x,\mu}+\bar k_{x,\mu}$, see \eqref{eq:k_kbar}, is even. In order to construct commuting expressions we insert an auxiliary Grassmann variable $\aux_{x,\mu}$ on each link using
\begin{equation}
  \label{eq:aux}
  \left(\int d\aux_{x,\mu}\,\aux_{x,\mu}\right)^{\F_{x,\mu}} = 1
\end{equation}
with the Grassmann parity
\begin{equation}
  \F_{x,\mu}\equiv
  \begin{cases}
    0 & \text{if } k_{x,\mu}+\bar k_{x,\mu} \text{ even}, \\
    1 & \text{if } k_{x,\mu}+\bar k_{x,\mu} \text{ odd}.
  \end{cases}
  \tag{\ref{eq:f}}
\end{equation}
For a given link, we insert \eqref{eq:aux} in front of \eqref{eq:Nested_Grass1} to obtain
\begin{align}
  \mathcal{G}_{x,\mu} &= \left(\int d\aux_{x,\mu}\aux_{x,\mu}\right)^{\F_{x,\mu}}
  \coprod_f\coprod_{a=1}^{k_{x,\mu}^f} \bar{\psi}_{x}^{f,i_{x,\mu}^{a+\kappa_{x,\mu}^f}}
    \prod_f\prod_{a=1}^{\bar k_{x,\mu}^f} \psi_{x}^{f,m_{x,\mu}^{a+\bar\kappa_{x,\mu}^f}}
    \coprod_f \coprod_{a=1}^{\bar k_{x,\mu}^f} \bar{\psi}_{x+\hat\mu}^{f,\ell_{x,\mu}^{a+\bar\kappa_{x,\mu}^f}}
    \prod_f\prod_{a=1}^{k_{x,\mu}^f} \psi_{x+\hat\mu}^{f,j_{x,\mu}^{a+\kappa_{x,\mu}^f}}\notag\\
  &=\left(\int\right)^{\!\F_{x,\mu}}
    \left[\left(\aux_{x,\mu}\right)^{\F_{x,\mu}}
    \coprod_f\coprod_{a=1}^{k_{x,\mu}^f} \bar{\psi}_{x}^{f,i_{x,\mu}^{a+\kappa_{x,\mu}^f}}
    \prod_f\prod_{a=1}^{\bar k_{x,\mu}^f} \psi_{x}^{f,m_{x,\mu}^{a+\bar\kappa_{x,\mu}^f}}
    \!\right]
    \left[\left(d\aux_{x,\mu}\right)^{\F_{x,\mu}}
    \coprod_f \coprod_{a=1}^{\bar k_{x,\mu}^f} \bar{\psi}_{x+\hat\mu}^{f,\ell_{x,\mu}^{a+\bar\kappa_{x,\mu}^f}}
    \prod_f\prod_{a=1}^{k_{x,\mu}^f} \psi_{x+\hat\mu}^{f,j_{x,\mu}^{a+\kappa_{x,\mu}^f}}
    \!\right],
    \label{eq:brackets}
\end{align}
where in the second step we have moved the differential of the auxiliary Grassmann variable in order to obtain commuting expressions. Owing to \eqref{eq:f}, no sign factors are produced in this step.

Next, we perform the product over links in \eqref{eq:Grassmann_block}. Since we now have commuting expressions, we can shift $(x,\mu)\to(x-\hat\mu,\mu)$ in the last bracket of \eqref{eq:brackets} to sort all original Grassmann variables according to their site,
\begin{align}\label{eq:Gint}
  \mathcal{G} = e^{\SM} \pif\prod_x\mathcal{G}_x\quad\text{with}\quad
  \mathcal{G}_x&=\prod_\mu
  \left[\left(\aux_{x,\mu}\right)^{\F_{x,\mu}}
    \coprod_f\coprod_{a=1}^{k_{x,\mu}^f} \bar{\psi}_{x}^{f,i_{x,\mu}^{a+\kappa_{x,\mu}^f}}\right]
    \left[\prod_f\prod_{a=1}^{\bar k_{x,\mu}^f} \psi_{x}^{f,m_{x,\mu}^{a+\bar\kappa_{x,\mu}^f}}\right]
    \\
    &\qquad\times
\left[\left(d\aux_{x,-\mu}\right)^{\F_{x,-\mu}}
    \coprod_f \coprod_{a=1}^{\bar k_{x,-\mu}^f} \bar{\psi}_{x}^{f,\ell_{x,-\mu}^{a+\bar\kappa_{x,-\mu}^f}}\right]
    \left[\prod_f\prod_{a=1}^{k_{x,-\mu}^f} \psi_{x}^{f,j_{x,-\mu}^{a+\kappa_{x,-\mu}^f}}\right],\notag
\end{align}
where we have defined
\begin{equation}
  \pif\equiv\prod_{x,\mu}\left(\int\right)^{\F_{x,\mu}}
  \tag{\ref{eq:pif}}
\end{equation}
to indicate the integration over the auxiliary Grassmann variables and where $(x,-\mu)$ again is a shorthand notation for $(x-\hat\mu,\mu)$.\footnote{To avoid potential confusion we point out that $\mathcal{G}_x\ne\prod_\mu\mathcal{G}_{x,\mu}$.}

We define the boundary conditions for the $\aux_{x,\mu}$ to be antiperiodic in time and periodic otherwise. The advantage of this definition is that the signs arising from the antiperiodic boundary conditions of $\psi$ and $\bar\psi$ in the step from \eqref{eq:brackets} to \eqref{eq:Gint} are absorbed through the antiperiodic boundary conditions of $\chi$ (or, more precisely, $d\chi$).

In \eqref{eq:Gint}, $\psi_x$ and $\bar\psi_x$ are not yet ordered in the integrand. In the next step we wish to gather all $\psi_x$ and $\bar\psi_x$ in $\mathcal{G}_x$. To do so, we first note that $\mathcal{G}_x$ can be written symbolically as $\prod_\mu A_\mu B_\mu C_\mu D_\mu$,
where $A_\mu$ and $C_\mu$ include $(\aux_{x,\mu})^{\F_{x,\mu}}$ and $(d\aux_{x,-\mu})^{\F_{x,-\mu}}$, respectively. Note that $A_\mu$ and $B_\mu$ have the same Grassmann parity, and so do $C_\mu$ and $D_\mu$. Thus both $A_\mu B_\mu$ and $C_\mu D_\mu$ are commuting. This allows us to write
\begin{align}
  \label{eq:ABCD}
  \prod_\mu (A_\mu B_\mu) (C_\mu D_\mu)
  &=\left[\prod_\mu A_\mu B_\mu\right]\left[ \prod_\mu C_\mu D_\mu\right]
  =\left[\coprod_\mu A_\mu \prod_\mu B_\mu\right] \left[\coprod_\mu C_\mu \prod_\mu D_\mu\right]\notag\\
  &=\coprod_\mu C_\mu \coprod_\mu A_\mu \prod_\mu B_\mu \prod_\mu D_\mu\,,
\end{align}
where in the first step we have distributed the product over $\mu$, in the second step we have applied \eqref{reorder} twice, and in the final step we have moved the first commuting bracket inside the second one. Using \eqref{eq:ABCD} in $\mathcal{G}_x$ yields
\begin{align}\label{eq:aux_all_Grassmanns_sorted_site}
  &\mathcal{G}_x = 
  \left[\coprod_\mu\left(d\aux_{x,-\mu}\right)^{\F_{x,-\mu}}\coprod_f \coprod_{a=1}^{\bar k_{x,-\mu}^f} \bar{\psi}_{x}^{f,\ell_{x,-\mu}^{a+\bar\kappa_{x,-\mu}^f}}\right]
    \left[\coprod_\mu\left(\aux_{x,\mu}\right)^{\F_{x,\mu}}\coprod_f\coprod_{a=1}^{k_{x,\mu}^f} \bar{\psi}_{x}^{f,i_{x,\mu}^{a+\kappa_{x,\mu}^f}}\right]
    \notag\\
  &\qquad\quad\times \left[\prod_\mu\prod_f\prod_{a=1}^{\bar k_{x,\mu}^f} \psi_{x}^{f,m_{x,\mu}^{a+\bar\kappa_{x,\mu}^f}}\right]
    \left[\prod_\mu\prod_f\prod_{a=1}^{k_{x,-\mu}^f} \psi_{x}^{f,j_{x,-\mu}^{a+\kappa_{x,-\mu}^f}}\right]
    \notag\\
  &= 
  \sigma_x^{(1)}K_x
    \left[\coprod_\mu\coprod_f\! \coprod_{a=1}^{\bar k_{x,-\mu}^f} \bar{\psi}_{x}^{f,\ell_{x,-\mu}^{a+\bar\kappa_{x,-\mu}^f}}\right]
    \left[\coprod_\mu\coprod_f\coprod_{a=1}^{k_{x,\mu}^f} \bar{\psi}_{x}^{f,i_{x,\mu}^{a+\kappa_{x,\mu}^f}}\right]
    \left[\prod_\mu\prod_f\prod_{a=1}^{\bar k_{x,\mu}^f} \psi_{x}^{f,m_{x,\mu}^{a+\bar\kappa_{x,\mu}^f}}\right]
    \left[\prod_\mu\prod_f\prod_{a=1}^{k_{x,-\mu}^f} \psi_{x}^{f,j_{x,-\mu}^{a+\kappa_{x,-\mu}^f}}\right]
\end{align}
with
\begin{equation}
  K_x= 
   \prod_\mu\left(\aux_{x,\mu}\right)^{\F_{x,\mu}}
   \coprod_\mu\left(d\aux_{x,-\mu}\right)^{\F_{x,-\mu}}
    \,. \tag{\ref{eq:Gxfx}}
\end{equation}
After the first step in \eqref{eq:aux_all_Grassmanns_sorted_site} all $\psi$ and all $\bar\psi$ are now gathered. In the second step we have moved the auxiliary Grassmann variables to the front and reordered them according to the prescription of \cite{Bloch:2022vqz} by commuting the Grassmann variables, which produces the local sign factor
\begin{align}
\sigma_x^{(1)} \equiv (-)^{p_1+p_2+p_3}\quad\text{with}\quad
p_1 &\equiv \sum_{\mu=1}^{d-1} \F_{x,-\mu}
\sum_{\nu=\mu+1}^d \bar k_{x,-\nu}\,,\notag\\
p_2 &\equiv \sum_{\mu=1}^{d-1} \F_{x,\mu} \sum_{\nu=\mu+1}^d (\F_{x,\nu}+k_{x,\nu})\,,\notag\\
p_3 &\equiv \sum_{\mu=1}^d \F_{x,\mu}\sum_{\nu=1}^d (\F_{x,-\nu}+\bar k_{x,-\nu})\,,
        \label{eq:sigma1}
\end{align}
where we used the sum \eqref{eq:k_kbar} over flavors for $k$ and $\bar k$.

\Cref{eq:Gint} still contains the mass term, which we now also Taylor-expand. Starting from \eqref{eq:mass} we write
\begin{align}
  \label{eq:SMftaylor}
  e^{\SM^f}=\prod_x\sum_{w_x^f}\frac{(2m_f)^{w_x^f}}{w_x^f!}(\bar\psi_x^f\psi_x^f)^{w_x^f}\,.
\end{align}
In principle, the sum over $w_x^f$ goes from zero to infinity. However, we will now show that the sum collapses to a single term once we integrate over all Grassmann variables. The point is that, due to the Grassmann integral, the only nonzero contributions to the partition function come from terms in which every Grassmann variable $\psi_x^{f,i}$ (and $\bar\psi_x^{f,i}$) occurs once. Let us fix the site $x$ and the flavor $f$ for the time being and define
\begin{equation}
  \hin_x \equiv \sum_\mu (k_{x,-\mu}^f+\bar k_{x,\mu}^f) \quad\text{and}\quad
  \hout_x \equiv \sum_\mu (\bar k_{x,-\mu}^f+k_{x,\mu}^f) \,,
  \tag{\ref{eq:hin_hout}}
\end{equation}
which is the number of incoming and outgoing hopping terms, respectively, at site $x$ for flavor $f$, see, for example, \eqref{eq:Grassmann_block}, where ``in'' and ``out'' correspond to $\psi$ and $\bar\psi$, respectively.
Since the total number of $\psi$ and $\bar\psi$ variables for fixed $x$ and $f$ equals $\Nc$ each,
we obtain the constraints
\begin{equation}
  \label{eq:constraint}
  w_x^f+\hin_x=\Nc=w_x^f+\hout_x \quad\text{and}\quad w_x^f\ge0\,.
\end{equation}
Hence, in the following we can, for each flavor, omit the sum over $w_x^f$ in \eqref{eq:SMftaylor}, replace $w_x^f$ by $\Nc-\hin_x$, and include a Heaviside step function $\Theta(w_x^f)$ as well as a Kronecker delta $\delta_{\hin_x,\hout_x}$.

We now use \eqref{reorder} to sort the $\psi$ and $\bar\psi$ in \eqref{eq:SMftaylor},
\begin{equation}
  \label{eq:1}
  (\bar\psi_x^f\psi_x^f)^{w_x^f}=\prod_{a=1}^{w_x^f}\bar\psi_{x}^{f,v_x^{f,a}}\psi_{x}^{f,v_x^{f,a}}
  =\coprod_{a=1}^{w_x^f}\bar\psi_{x}^{f,v_x^{f,a}}\prod_{a=1}^{w_x^f}\psi_{x}^{f,v_x^{f,a}}\,,
\end{equation}
where the Einstein summation convention is used for the color indices $v_x^{f,a}$. We then perform the product over flavors and again use \eqref{reorder} to obtain
\begin{equation}\label{eq:Mass_Grassmanns}
  \prod_f\coprod_{a=1}^{w_x^f}\bar\psi_{x}^{f,v_x^{f,a}}\prod_{a=1}^{w_x^f}\psi_{x}^{f,v_x^{f,a}}
  =\coprod_f\coprod_{a=1}^{w_x^f}\bar\psi_{x}^{f,v_x^{f,a}}\prod_f\prod_{a=1}^{w_x^f}\psi_{x}^{f,v_x^{f,a}}\,.
\end{equation}
We now substitute \eqref{eq:SMftaylor}, multiplied over flavors, into \eqref{eq:Gint} (using \eqref{eq:aux_all_Grassmanns_sorted_site} for $\mathcal{G}_x$ and \eqref{eq:Mass_Grassmanns}) in such a way that $\psi$ and $\bar\psi$ are not mixed. We then split off the part that only contains the original Grassmann variables, which is
\begin{align}
  \label{eq:Psix}
  \Psi
  &\equiv\prod_x\left[\coprod_\mu\coprod_f \coprod_{a=1}^{\bar k_{x,-\mu}^f} \bar{\psi}_{x}^{f,\ell_{x,-\mu}^{a+\bar\kappa_{x,-\mu}^f}}\right]
  \left[\coprod_\mu\coprod_f\coprod_{a=1}^{k_{x,\mu}^f} \bar{\psi}_{x}^{f,i_{x,\mu}^{a+\kappa_{x,\mu}^f}}\right]
  \left[\coprod_f\coprod_{a=1}^{w_x^f}\bar\psi_{x}^{f,v_x^{f,a}}\right]\notag\\
  & \qquad\times\left[\prod_f\prod_{a=1}^{w_x^f}\psi_{x}^{f,v_x^{f,a}}\right]
  \left[\prod_\mu\prod_f\prod_{a=1}^{\bar k_{x,\mu}^f} \psi_{x}^{f,m_{x,\mu}^{a+\bar\kappa_{x,\mu}^f}}\right]
  \left[\prod_\mu\prod_f\prod_{a=1}^{k_{x,-\mu}^f} \psi_{x}^{f,j_{x,-\mu}^{a+\kappa_{x,-\mu}^f}}\right].
\end{align}
To prepare for the integration over the original Grassmann variables it is useful to change the order of the products over $\mu$ and $f$ in each of the factors in \eqref{eq:Psix} so that the Grassmann variables are sorted by flavor. We obtain
\begin{align}
  \Psi
  & = \prod_x
    \left[\coprod_f\coprod_\mu \coprod_{a=1}^{\bar k_{x,-\mu}^f} \bar{\psi}_{x}^{f,\ell_{x,-\mu}^{a+\bar\kappa_{x,-\mu}^f}}\right]
    \left[\coprod_f\coprod_\mu\coprod_{a=1}^{k_{x,\mu}^f} \bar{\psi}_{x}^{f,i_{x,\mu}^{a+\kappa_{x,\mu}^f}}\right]
    \left[\coprod_f\coprod_{a=1}^{w_x^f}\bar\psi_{x}^{f,v_x^{f,a}}\right]\notag\\
  & \qquad\times \left[\prod_f\prod_{a=1}^{w_x^f}\psi_{x}^{f,v_x^{f,a}}\right]
  \left[\prod_f\prod_\mu\prod_{a=1}^{\bar k_{x,\mu}^f} \psi_{x}^{f,m_{x,\mu}^{a+\bar\kappa_{x,\mu}^f}}\right]
    \left[\prod_f\prod_\mu\prod_{a=1}^{k_{x,-\mu}^f} \psi_{x}^{f,j_{x,-\mu}^{a+\kappa_{x,-\mu}^f}}\right] .
    \label{eq:Psixf}
\end{align}
All sign factors that are generated by changing the order of the products cancel after the product over sites is performed.
To show this, we first consider the two brackets involving $k_{x,\mu}^f$ in \eqref{eq:Psix}, which can symbolically be written as
\begin{equation}
  \coprod_\mu\coprod_f (\alpha_{\mu}^{f})^{k_{\mu}^f} \quad\text{and}\quad
  \prod_\mu\prod_f (\beta_{\mu}^{f})^{k_{\mu}^f}\,,
\end{equation}
where we use $(\alpha)^k$ for a coproduct and $(\beta)^k$ for a product of $k$ different Grassmann variables (the Grassmann variables inside these products do not get reordered).
When reversing the order of the products over $\mu$ and $f$, the commutations of Grassmann variables generate a sign factor $\sigma$ that is identical for both terms since the same commutations are performed but in opposite directions,
\begin{subequations}
\begin{align}
  \coprod_\mu\coprod_f (\alpha_{\mu}^{f})^{k_{\mu}^f}&= (\alpha_d^{{\Nf}})^{k_d^{\Nf}} \cdots (\alpha_d^{1})^{k_d^1} \cdots
  (\alpha_1^{{\Nf}})^{k_1^{\Nf}} \cdots (\alpha_1^{1})^{k_1^1}\notag\\
  &= \sigma \times (\alpha_d^{{\Nf}})^{k_d^{\Nf}}\cdots (\alpha_1^{{\Nf}})^{k_1^{\Nf}}
  \cdots(\alpha_d^{1})^{k_d^1} \cdots (\alpha_1^{1})^{k_1^1}
  = \sigma\coprod_f \coprod_\mu (\alpha_{\mu}^{f})^{k_{\mu}^f}\,,\\
  \prod_\mu\prod_f (\beta_{\mu}^{f})^{k_{\mu}^f}&=(\beta_1^{1})^{k_1^1} \cdots (\beta_1^{{\Nf}})^{k_1^{\Nf}}\cdots
  (\beta_d^{1})^{k_d^1} \cdots (\beta_d^{{\Nf}})^{k_d^{\Nf}}\notag\\
  &=\sigma \times (\beta_1^{1})^{k_1^1} \cdots (\beta_d^{1})^{k_d^1} \cdots
  (\beta_1^{{\Nf}})^{k_1^{\Nf}} \cdots (\beta_d^{{\Nf}})^{k_d^{\Nf}}
    =\sigma\prod_f\prod_\mu (\beta_{\mu}^{f})^{k_{\mu}^f}\,.
\end{align}
\end{subequations}
Hence, since $\sigma^2=1$, no net sign factor is generated when considering the product of both terms.
A similar argument applies to the two brackets involving $\bar k^f_{x,\mu}$ in \eqref{eq:Psix}.

In \eqref{eq:Psixf} we now merge three products over $f$ into one, for both $\bar\psi$ and $\psi$, resulting in
\begin{equation}
  \label{eq:Psi}
  \Psi = \prod_x\sigma_x^{(2)}\coprod_f\bar\Psi_x^f\prod_f\Psi_x^f
\end{equation}
with
\begin{subequations}
\begin{align}
  \bar\Psi_x^f&=
    \left[\coprod_\mu \coprod_{a=1}^{\bar k_{x,-\mu}^f} \bar{\psi}_{x}^{f,\ell_{x,-\mu}^{a+\bar\kappa_{x,-\mu}^f}}\right]
    \left[\coprod_\mu\coprod_{a=1}^{k_{x,\mu}^f} \bar{\psi}_{x}^{f,i_{x,\mu}^{a+\kappa_{x,\mu}^f}}\right]
    \left[\coprod_{a=1}^{w_x^f}\bar\psi_{x}^{f,v_x^{f,a}}\right],\\
  \Psi_x^f &= \left[\prod_{a=1}^{w_x^f}\psi_{x}^{f,v_x^{f,a}}\right]
  \left[\prod_\mu\prod_{a=1}^{\bar k_{x,\mu}^f} \psi_{x}^{f,m_{x,\mu}^{a+\bar\kappa_{x,\mu}^f}}\right]
    \left[\prod_\mu\prod_{a=1}^{k_{x,-\mu}^f} \psi_{x}^{f,j_{x,-\mu}^{a+\kappa_{x,-\mu}^f}}\right]
\end{align}
\end{subequations}
and a new sign factor
\begin{equation}
  \sigma_x^{(2)} \equiv (-)^{p_4+p_5} \quad\text{with}\quad
  p_4 \equiv \sum_{f=1}^{\Nf-1} k_x^{f+}\sum_{f'=f+1}^{\Nf} w_x^{f'}
        + \sum_{f=1}^{\Nf-1} \bark_x^{f-} \sum_{f'=f+1}^{\Nf} \left(w_x^{f'}+k_x^{f'+}\right)
        \quad\text{and}\quad p_5\equiv p_4(k\leftrightarrow\bar k) \,,
  \label{eq:sigma2}
\end{equation}
where
\begin{equation}
  \label{eq:2}
  k_x^{f\pm}\equiv\sum_\mu k_{x,\pm\mu}^f\quad\text{and}\quad
  \bar k_x^{f\pm}\equiv\sum_\mu \bar k_{x,\pm\mu}^f\,.
\end{equation}
Since $\bar\Psi_x^f$ and $\Psi_x^f$ have the same Grassmann parity due to \eqref{eq:constraint}, we can use \eqref{reorder} to rewrite \eqref{eq:Psi} as
\begin{align}
  \Psi = \prod_x\sigma_x^{(2)}\prod_f\bar\Psi_x^f \Psi_x^f\,.
    \label{eq:Psixf2}
\end{align}

We can now integrate out the original Grassmann variables on each site flavor by flavor. For a single flavor on a single site we have
\begin{align}
  \int d \psi d\bar\psi
  \coprod_{a=1}^{A}\bar\psi^{i^a}
  \prod_{b=1}^{B}\psi^{j^b}
  &=\int 
    \coprod_{k=1}^{\Nc}d\psi^k 
    \prod_{\ell=1}^{\Nc}d\bar\psi^\ell
    \coprod_{a=1}^{A}\bar\psi^{i^a}
    \prod_{b=1}^{B}\psi^{j^b}
  =
  \delta_{A,\Nc}\varepsilon_{i_1\dotsb i_A}\cdot
  \delta_{B,\Nc}\varepsilon_{j_1\dotsb j_B}\,,
    \label{eq:int_out}
\end{align}
where we have omitted the site and flavor indices for simplicity.
To apply this integral to \eqref{eq:Psixf2} we move each commuting pair of differentials $d\bar\psi_x^f d\psi_x^f$ in \eqref{eq:Gint0} in front of the corresponding $\bar\Psi_x^f\Psi_x^f$. After integrating $\bar\psi_x^f$ and $\psi_x^f$ for every flavor $f$ and every site $x$ we obtain
\begin{equation}
  \int \left[\prod_x d\psi_x d\bar\psi_x\right] \mathcal{G}
  = \pif \prod_x K_x\sigma_x\prod_f\frac{(2m_f)^{w_x^f}}{w_x^f!} E_x^f \Theta(w_x^f) \, \delta_{\hin_x,\hout_x}  \,,
\tag{\ref{eq:Grassmann_evaluation}}
\end{equation}
where
\begin{equation}
  \label{eq:sigma}
  \sigma_x=\sigma_x^{(1)}\sigma_x^{(2)}
\end{equation}
is the combined sign factor on site $x$ and
\begin{align}
  E_x^f&=
      \epsilon_{v_x^{f,1},\dots,v_x^{f,w_x^f},
      i_{x,1}^{\kappa_{x,1}^f+1},\dots,i_{x,1}^{\kappa_{x,1}^f+k_{x,1}^f}, \dots,
      i_{x,d}^{\kappa_{x,d}^f+1},\dots,i_{x,d}^{\kappa_{x,d}^f+k_{x,d}^f},
      \ell_{x,-1}^{\bar\kappa_{x,-1}^f+1},\dots,\ell_{x,-1}^{\bar\kappa_{x,-1}^f+\bar k_{x,-1}^f}, \dots,
      \ell_{x,-d}^{\bar\kappa_{x,-d}^f+1},\dots,\ell_{x,-d}^{\bar\kappa_{x,-d}^f+\bar k_{x,-d}^f}}\notag\\
       &\quad\times  \epsilon_{v_x^{f,1},\dots,v_x^{f,w_x^f},
      m_{x,1}^{\bar\kappa_{x,1}^f+1},\dots,m_{x,1}^{\bar\kappa_{x,1}^f+\bar k_{x,1}^f}, \dots,
      m_{x,d}^{\bar\kappa_{x,d}^f+1},\dots,m_{x,d}^{\bar\kappa_{x,d}^f+\bar k_{x,d}^f},
      j_{x,-1}^{\kappa_{x,-1}^f+1},\dots,j_{x,-1}^{\kappa_{x,-1}^f+k_{x,-1}^f}, \dots,
         j_{x,-d}^{\kappa_{x,-d}^f+1},\dots,j_{x,-d}^{\kappa_{x,-d}^f+k_{x,-d}^f}}\,,
         \tag{\ref{eq:Grassmann_epsilon_site}}
\end{align}
is the result of the integration over the original Grassmann variables for flavor $f$.

\section{\boldmath Expansion coefficients for a $2\times 2$ lattice}
\label{app:Zn}

The expansion coefficients $Z_n$ in \eqref{eq:Zexp} can be computed analytically for a $2\times 2$ lattice and are given, up to $n=4$, by
\begin{subequations}
\begin{align}
	Z_0(m,\mu) 	
	&= 4096 m^{12} + 24576 m^{10} + 53248 m^{8} + 50944 m^{6} + 21248 m^{4} + \frac{9760 m^{2}}{3} + \frac{998}{9}
	\notag\\	
	&\quad+ \left(256 m^{6} + 384 m^{4} + 160 m^{2} + 16\right) \cosh{\left(6 \mu \right)} + 2 \cosh{\left(12 \mu \right)} \,,
\\
	Z_1(m,\mu) &= \frac{1024 m^{8}}{3} + \frac{10240 m^{6}}{9} + \frac{29632 m^{4}}{27} + \frac{26080 m^{2}}{81} + \frac{5380}{243} + \left(\frac{128 m^{4}}{3} + \frac{416 m^{2}}{9} + \frac{80}{9}\right) \cosh{\left(6 \mu \right)} \,,
\\
	Z_2(m,\mu) 	
	&= \frac{4096 m^{12}}{9} + \frac{8192 m^{10}}{3} + \frac{482048 m^{8}}{81} + \frac{1405312 m^{6}}{243} + \frac{1820704 m^{4}}{729} + \frac{100624 m^{2}}{243} + \frac{13034}{729}
	\notag\\	
	&\quad+ \left(\frac{2432 m^{6}}{81} + \frac{11936 m^{4}}{243} + \frac{17648 m^{2}}{729} + \frac{2344}{729}\right) \cosh{\left(6 \mu \right)} + \frac{2}{9} \cosh{\left(12 \mu \right)} \,,
\\
	Z_3(m,\mu) 	
	&= \frac{2048 m^{12}}{81} + \frac{4096 m^{10}}{27} + \frac{804352 m^{8}}{2187} + \frac{325312 m^{6}}{729} + \frac{21232 m^{4}}{81} + \frac{134452 m^{2}}{2187} + \frac{8887}{2187} 
	\notag\\	
	&\quad+ \left(\frac{448 m^{6}}{243} + \frac{16768 m^{4}}{2187} + \frac{14548 m^{2}}{2187} + \frac{2714}{2187}\right) \cosh{\left(6 \mu \right)} + \frac{1}{81} \cosh{\left(12 \mu \right)} \,,
\\
	Z_4(m,\mu) &= \frac{166400 m^{12}}{6561} + \frac{332800 m^{10}}{2187} + \frac{2201008 m^{8}}{6561} + \frac{2201512 m^{6}}{6561} + \frac{1788793 m^{4}}{11664} + \frac{2968507 m^{2}}{104976} + \frac{646769}{419904} 
	\notag\\
	&\quad+ \left(\frac{3880 m^{6}}{2187} + \frac{22054 m^{4}}{6561} + \frac{13085 m^{2}}{6561} + \frac{938}{2916}\right) \cosh{\left(6 \mu \right)} + \frac{325}{26244} \cosh{\left(12 \mu \right)} \,.
\end{align}
\end{subequations}

\bibliographystyle{elsarticle-num}
\bibliography{qcd_sce} 

\begin{thebibliography}{10}
\expandafter\ifx\csname url\endcsname\relax
  \def\url#1{\texttt{#1}}\fi
\expandafter\ifx\csname urlprefix\endcsname\relax\def\urlprefix{URL }\fi
\expandafter\ifx\csname href\endcsname\relax
  \def\href#1#2{#2} \def\path#1{#1}\fi

\bibitem{deForcrand:2010ys}
P.~de~Forcrand, Simulating {QCD} at finite density, PoS LAT2009 (2009) 010.
\newblock \href {http://arxiv.org/abs/1005.0539} {\path{arXiv:1005.0539}},
  \href {http://dx.doi.org/10.22323/1.091.0010}
  {\path{doi:10.22323/1.091.0010}}.

\bibitem{Aarts:2015tyj}
G.~Aarts, Introductory lectures on lattice {QCD} at nonzero baryon number, J.
  Phys. Conf. Ser. 706~(2) (2016) 022004.
\newblock \href {http://arxiv.org/abs/1512.05145} {\path{arXiv:1512.05145}},
  \href {http://dx.doi.org/10.1088/1742-6596/706/2/022004}
  {\path{doi:10.1088/1742-6596/706/2/022004}}.

\bibitem{Rossi:1984cv}
P.~Rossi, U.~Wolff, Lattice {QCD} with fermions at strong coupling: A dimer
  system, Nucl. Phys. B 248 (1984) 105.
\newblock \href {http://dx.doi.org/10.1016/0550-3213(84)90589-3}
  {\path{doi:10.1016/0550-3213(84)90589-3}}.

\bibitem{Karsch:1988zx}
F.~Karsch, K.-H. Mütter, Strong coupling {QCD} at finite baryon number
  density, Nucl. Phys. B 313 (1989) 541.
\newblock \href {http://dx.doi.org/10.1016/0550-3213(89)90396-9}
  {\path{doi:10.1016/0550-3213(89)90396-9}}.

\bibitem{Fromm:2010lga}
M.~Fromm, Lattice {QCD} at strong coupling: thermodynamics and nuclear physics,
  Ph.D. thesis, ETH Zürich (2010).
\newblock \href {http://dx.doi.org/10.3929/ETHZ-A-006414247}
  {\path{doi:10.3929/ETHZ-A-006414247}}.

\bibitem{deForcrand:2009dh}
P.~de~Forcrand, M.~Fromm, Nuclear physics from lattice {QCD} at strong
  coupling, Phys. Rev. Lett. 104 (2010) 112005.
\newblock \href {http://arxiv.org/abs/0907.1915} {\path{arXiv:0907.1915}},
  \href {http://dx.doi.org/10.1103/PhysRevLett.104.112005}
  {\path{doi:10.1103/PhysRevLett.104.112005}}.

\bibitem{Forcrand2014}
P.~de~Forcrand, J.~Langelage, O.~Philipsen, W.~Unger, Lattice {QCD} phase
  diagram in and away from the strong coupling limit, Phys. Rev. Lett. 113
  (2014) 152002.
\newblock \href {http://arxiv.org/abs/1406.4397} {\path{arXiv:1406.4397}},
  \href {http://dx.doi.org/10.1103/PhysRevLett.113.152002}
  {\path{doi:10.1103/PhysRevLett.113.152002}}.

\bibitem{Bilic:1991qy}
N.~Bilic, K.~Demeterfi, B.~Petersson, Strong coupling analysis of the chiral
  phase transition at finite chemical potential and finite temperature, Nucl.
  Phys. B 377 (1992) 651.
\newblock \href {http://dx.doi.org/10.1016/0550-3213(92)90305-U}
  {\path{doi:10.1016/0550-3213(92)90305-U}}.

\bibitem{Marchis:2017oqi}
C.~Marchis, C.~Gattringer, Dual representation of lattice {QCD} with worldlines
  and worldsheets of {Abelian} color fluxes, Phys. Rev. D 97 (2018) 034508.
\newblock \href {http://arxiv.org/abs/1712.07546} {\path{arXiv:1712.07546}},
  \href {http://dx.doi.org/10.1103/PhysRevD.97.034508}
  {\path{doi:10.1103/PhysRevD.97.034508}}.

\bibitem{Gagliardi:2019cpa}
G.~Gagliardi, W.~Unger, New dual representation for staggered lattice {QCD},
  Phys. Rev. D 101 (2020) 034509.
\newblock \href {http://arxiv.org/abs/1911.08389} {\path{arXiv:1911.08389}},
  \href {http://dx.doi.org/10.1103/PhysRevD.101.034509}
  {\path{doi:10.1103/PhysRevD.101.034509}}.

\bibitem{Unger:2025sjh}
W.~Unger, J.~Kim, P.~Pattanaik, The chiral critical point from the strong
  coupling expansion, PoS LATTICE2024 (2025) 166.
\newblock \href {http://arxiv.org/abs/2502.06679} {\path{arXiv:2502.06679}},
  \href {http://dx.doi.org/10.22323/1.466.0166}
  {\path{doi:10.22323/1.466.0166}}.

\bibitem{Prokofiev:2001zz}
N.~Prokof'ev, B.~Svistunov, Worm algorithms for classical statistical models,
  Phys. Rev. Lett. 87 (2001) 160601.
\newblock \href {http://dx.doi.org/10.1103/PhysRevLett.87.160601}
  {\path{doi:10.1103/PhysRevLett.87.160601}}.

\bibitem{Levin:2006jai}
M.~Levin, C.~P. Nave, Tensor renormalization group approach to two-dimensional
  classical lattice models, Phys. Rev. Lett. 99 (2007) 120601.
\newblock \href {http://arxiv.org/abs/cond-mat/0611687}
  {\path{arXiv:cond-mat/0611687}}, \href
  {http://dx.doi.org/10.1103/PhysRevLett.99.120601}
  {\path{doi:10.1103/PhysRevLett.99.120601}}.

\bibitem{Xie_2012}
Z.~Y. Xie, J.~Chen, M.~P. Qin, J.~W. Zhu, L.~P. Yang, T.~Xiang, Coarse-graining
  renormalization by higher-order singular value decomposition, Phys. Rev. B 86
  (2012) 045139.
\newblock \href {http://dx.doi.org/10.1103/physrevb.86.045139}
  {\path{doi:10.1103/physrevb.86.045139}}.

\bibitem{DeLathauwer2000}
L.~De~Lathauwer, B.~De~Moor, J.~Vandewalle, A multilinear singular value
  decomposition, SIAM Journal on Matrix Analysis and Applications 21 (2000)
  1253.
\newblock \href {http://dx.doi.org/10.1137/S0895479896305696}
  {\path{doi:10.1137/S0895479896305696}}.

\bibitem{Bloch:2021mjw}
J.~Bloch, R.~G. Jha, R.~Lohmayer, M.~Meister, Tensor renormalization group
  study of the three-dimensional {$O(2)$} model, Phys. Rev. D 104 (2021)
  094517.
\newblock \href {http://arxiv.org/abs/2105.08066} {\path{arXiv:2105.08066}},
  \href {http://dx.doi.org/10.1103/PhysRevD.104.094517}
  {\path{doi:10.1103/PhysRevD.104.094517}}.

\bibitem{Gu:2010yh}
Z.-C. Gu, F.~Verstraete, X.-G. Wen, Grassmann tensor network states and its
  renormalization for strongly correlated fermionic and bosonic states (2010).
\newblock \href {http://arxiv.org/abs/1004.2563} {\path{arXiv:1004.2563}},
  \href {http://dx.doi.org/10.48550/arXiv.1004.2563}
  {\path{doi:10.48550/arXiv.1004.2563}}.

\bibitem{Shimizu:2014uva}
Y.~Shimizu, Y.~Kuramashi, Grassmann tensor renormalization group approach to
  one-flavor lattice {Schwinger} model, Phys. Rev. D 90 (2014) 014508.
\newblock \href {http://arxiv.org/abs/1403.0642} {\path{arXiv:1403.0642}},
  \href {http://dx.doi.org/10.1103/PhysRevD.90.014508}
  {\path{doi:10.1103/PhysRevD.90.014508}}.

\bibitem{Takeda:2014vwa}
S.~Takeda, Y.~Yoshimura, Grassmann tensor renormalization group for the
  one-flavor lattice {Gross-Neveu} model with finite chemical potential, PTEP
  2015 (2015) 043B01.
\newblock \href {http://arxiv.org/abs/1412.7855} {\path{arXiv:1412.7855}},
  \href {http://dx.doi.org/10.1093/ptep/ptv022}
  {\path{doi:10.1093/ptep/ptv022}}.

\bibitem{Sakai:2017jwp}
R.~Sakai, S.~Takeda, Y.~Yoshimura, Higher order tensor renormalization group
  for relativistic fermion systems, PTEP 2017 (2017) 063B07.
\newblock \href {http://arxiv.org/abs/1705.07764} {\path{arXiv:1705.07764}},
  \href {http://dx.doi.org/10.1093/ptep/ptx080}
  {\path{doi:10.1093/ptep/ptx080}}.

\bibitem{Meurice:2020pxc}
Y.~Meurice, R.~Sakai, J.~Unmuth-Yockey, Tensor lattice field theory for
  renormalization and quantum computing, Rev. Mod. Phys. 94 (2022) 025005.
\newblock \href {http://arxiv.org/abs/2010.06539} {\path{arXiv:2010.06539}},
  \href {http://dx.doi.org/10.1103/RevModPhys.94.025005}
  {\path{doi:10.1103/RevModPhys.94.025005}}.

\bibitem{Bloch:2022vqz}
J.~Bloch, R.~Lohmayer, Grassmann higher-order tensor renormalization group
  approach for two-dimensional strong-coupling {QCD}, Nucl. Phys. B 986 (2023)
  116032.
\newblock \href {http://arxiv.org/abs/2206.00545} {\path{arXiv:2206.00545}},
  \href {http://dx.doi.org/10.1016/j.nuclphysb.2022.116032}
  {\path{doi:10.1016/j.nuclphysb.2022.116032}}.

\bibitem{Milde2023}
P.~Milde, Tensor networks for four-dimensional strong-coupling {QCD}, Master's
  thesis, University of Regensburg (2023).

\bibitem{Creutz:1978ub}
M.~Creutz, On invariant integration over {SU(N)}, J. Math. Phys. 19 (1978)
  2043.
\newblock \href {http://dx.doi.org/10.1063/1.523581}
  {\path{doi:10.1063/1.523581}}.

\bibitem{Gagliardi:2018tkz}
G.~Gagliardi, W.~Unger, Towards a dual representation of lattice {QCD}, PoS
  LATTICE2018 (2018) 224.
\newblock \href {http://arxiv.org/abs/1811.02817} {\path{arXiv:1811.02817}},
  \href {http://dx.doi.org/10.22323/1.334.0224}
  {\path{doi:10.22323/1.334.0224}}.

\bibitem{Borisenko:2018csw}
O.~Borisenko, S.~Voloshyn, V.~Chelnokov, {SU(N)} polynomial integrals and some
  applications, Rept. Math. Phys. 85 (2020) 129.
\newblock \href {http://arxiv.org/abs/1812.06069} {\path{arXiv:1812.06069}},
  \href {http://dx.doi.org/10.1016/S0034-4877(20)30015-X}
  {\path{doi:10.1016/S0034-4877(20)30015-X}}.

\bibitem{Gattringer.2010}
C.~Gattringer, C.~B. Lang, Quantum Chromodynamics on the Lattice: An
  Introductory Presentation, Lecture Notes in Physics, {Springer Berlin
  Heidelberg}, 2010.
\newblock \href {http://dx.doi.org/10.1007/978-3-642-01850-3}
  {\path{doi:10.1007/978-3-642-01850-3}}.

\bibitem{10.7717/peerj-cs.103}
{A. Meurer et al.}, Sympy: symbolic computing in {Python}, PeerJ Computer
  Science 3 (2017) e103.
\newblock \href {http://dx.doi.org/10.7717/peerj-cs.103}
  {\path{doi:10.7717/peerj-cs.103}}.

\end{thebibliography}

\end{document}